\documentclass[a4paper,final]{amsart}
\usepackage[utf8]{inputenc}
\usepackage[T1]{fontenc}
\usepackage{bm}
\usepackage{graphicx}
\usepackage[normalem]{ulem}
\usepackage[foot]{amsaddr}
\usepackage{amsmath}
\usepackage{amssymb}
\usepackage{natbib}
\usepackage[margin=3cm]{geometry}
\usepackage{pgfplots}
\pgfplotsset{compat=1.18}
\usepgfplotslibrary{statistics}
\usepackage{xcolor}
\definecolor{prussian}{HTML}{1E8765}
\usepackage{hyperref}
\hypersetup{colorlinks=true,allcolors=prussian}
\usepackage{salmon}
\hypersetup{
 pdfauthor={William Waites, Philip Gillibrand, Thomas Adams, Rek Bell, Duncan Guthrie, Crawford Revie, Meadhbh Moriarty},
 pdftitle={Infection Pressure on Fish in Cages},
 pdfkeywords={},
 pdfsubject={},
 pdflang={English}}
\usepackage{draftwatermark}
\SetWatermarkColor{red!20}
\SetWatermarkScale{2}
\begin{document}
\date{\today}
\title{Infection Pressure on Fish in Cages}
\author[W. Waites]{William Waites$^{1,2}$}
\address{$^{1}$ IT Innovation Centre, Electronics and Computer Science, University of Southampton}
\address{$^{2}$ Caledonian Maritime Research CIC, Edinburgh}
\email{w.waites@soton.ac.uk}
\author[P. Gillibrand]{Philip Gillibrand$^{3}$}
\address{$^{3}$ Mowi Scotland Ltd, Fort William}
\author[T. Adams]{Thomas Adams$^{4}$}
\address{$^{4}$ Scottish Sea Farms Ltd, Stirling}
\author[R. Bell]{Rek Bell$^5$}
\address{$^5$ kokorobot.ca, North Saanich}
\author[D. Guthrie]{Duncan Guthrie$^{6}$}
\address{$^{6}$ Computer and Information Sciences, University of Strathclyde, Glasgow}
\author[T. Kragesteen]{Tróndur Kragesteen$^{7}$}
\address{$^{7}$ Firum, Aquaculture research station, Faroe Island, Hósvík}
\author[C. Revie]{Crawford Revie$^{6}$}
\author[M. Moriarty]{Meadhbh Moriarty$^{8}$}
\address{$^{8}$ Marine Directorate, Scottish Government, Aberdeen}
\begin{abstract}
  We address the question of how to connect predictions by hydrodynamic models of how sea lice move in water to observable measures that count the number of lice on each fish in a cage in the water. This question is important for management and regulation of aquacultural practice that tries to maximise food production and minimise risk to the environment. We do this through a simple rule-based model of interaction between sea lice and caged fish. The model is simple: sea lice can attach and detach from a fish. The model has a novel feature, encoding what is known as a \emph{master equation} producing a time-series of distributions of lice on fish that one might expect to find if a cage full of fish were placed at any given location. To demonstrate how this works, and to arrive at a rough estimate of the interaction rates, we fit a simplified version of the model with three free parameters to publicly available data about an experiment with sentinel cages in Loch Linnhe in Scotland. Our construction, coupled to the hydrodynamic models driven by surveillance data from industrial farms,  quantifies the environmental impact as: what would the infection burden look like in a notional cage at any location and how does it change with time?
\end{abstract}
\maketitle

\section{Introduction}
\label{sec:intro}

Salmonids are vulnerable to infection by parasitic copepods, especially salmon lice (\emph{Lepeophtheirus salmonis}). This is a problem for farmed salmon, an important food stock in Northern Europe, Canada and Chile where the fish are kept in cages immersed in open water. To quantify the risk and take measures to avoid transmission of infection between cages and to wild fish, it is necessary to understand how the planktonic larval stages of the salmon lice  move in open water, and the dynamics of the processes by which they attach to and infest host fish.

Salmon lice have several stages in their lifecycle. In open water, the larval stages are planktonic, that is, they primarily move with the water, carried on currents and tides. These copepodid larvae have a secondary motion in the vertical direction as they rise or sink in response to light or salinity\citep{bron_settlement_1991,heuch_experimental_1995,heuch_diel_1995,flamarique_ontogenetic_2000,johnsen_vertical_2014}. When such a larva encounters a fish, it may attach to it and begin the next stage of its lifecycle, culminating in adult stages that release the next generation of larvae into the water column~\citep{moriarty_gap_2024}. Once attached, in principle, the louse may also detach.

Mathematical models of these processes are generally formulated in the following way. Given a particle (representing a planktonic stage salmon louse) in a particular volume of seawater $V$ in the neighbourhood of a position $x$ at time $t_0$, and given the bathymetry in the area taken together with tides and currents, what is the chance of finding that particle in some volume of water $V^\prime$ in the neighbourhood of position $x^\prime$ at time $t > t_0$? This can easily be generalised to accommodate the idea of multiple sources and gives rise to a time-varying density field, $\rho(x, t)$ tells us the chance of finding a copepod at any position at any time.

The density model for such an application is typically derived using a combination of a hydrodynamic model (simulating the Navier-Stokes equation of fluid dynamics with suitable boundary conditions) and particle tracking~\citep{johnsen_vertical_2014,moriarty_gap_2024}. However, many constants relate to quantities that cannot be derived from first principles such as the drag of the copepods as they are carried along, or their degree of self-motility. Calibrating these models is a challenge as is checking the correctness of the calibration.

To validate these models and calibrate them correctly, we must compare to empirical data. The sea lice lifecycle means that they occur as either planktonic larvae or as attached ectoparasites on farmed, wild or sentinel fish host populations. Surveillance methodologies and strategies for each of these stages should be optimized to obtain data that meet the specific needs of control. \citet{rabe_searching_2024}, provide a review of surveillance methodologies,  advocating using modelling as the most effective way of using surveillance data to inform subsequent model improvement in an adaptive feedback loop. Here we use data from sentinel cages data. Smaller than the cages used in farming, these cages are deployed in known locations for a specific amount of time (typically one or two weeks) and stocked with a standard number of fish~\citep{pert_using_2014}. At the end of a deployment, the fish are examined and their lice counted. We concern ourselves mainly with these data, but the method that we elaborate here could perhaps be extended to the salmon farm cage data; weekly sea lice counts, or wild salmon data; form catching fish in open water and examining them, counting the number of sea lice that have infected them~\citep{middlemas_relationship_2013}.

The important question thence arises of how to compare these two quantities: the expected density of salmon lice in the vicinity of a sentinel cage as a function of time, and the distribution of invective burden on the fish in that cage at the end of the deployment. Clearly, to do this correctly, some account of how free salmon lice come to attach themselves to the fish is needed. In this paper, we give a \emph{semi-mechanistic} account of this process.

We call it ``semi-mechanistic'' because our argument begins from the fact that the parasites attach themselves to and detach themselves from fish at some characteristic rate or rates, but we do not give an account of the mechanism through which this takes place. Nor do we provide a detailed picture of the fishes' immune response which presumably affects these rates. Nevertheless, the model we present is able to capture, at a phenomenological level, the typical long tailed distribution of sea lice counts from a simplified biological argument that only requires, as input, the ambient density of salmon lice.

\section{A Master Equation for Copepod-Salmonid Interaction}
\label{sec:master}

The mathematical construction that we use here is similar to what we used in a previous study about heterogeneity of viral loads in COVID-19~\citep{waites_compositional_2022}. More fundamentally, it uses a convenient encoding of a description of a dynamical system to produce and solve a \emph{master equation}~\citep{nordsieck_theory_1940} that describes the time evolution of the distribution of infection burden (lice count) that we can expect to see on the fish in a sentinel cage. It is important to recognise that this formulation does not describe a trajectory of the system where, if the system is small enough for stochastic effects to matter (as indeed it is) we would sample many trajectories in order to approximate the distribution of trajectories. Rather, we do the reverse; we formulate the problem so that we can directly get a time-varying trajectory of distributions for finding any particular number of lice on any given fish.

We describe the model using a \emph{rule-based} language~\citep{danos_formal_2004}. This formulation allows for stochastic simulation as well as derivation of the corresponding differential equations for the large population average of observable quantities.

\subsection{Attachment of ectoparasites to fish}
\label{sec:org4a41eca}

Inside the cage, we assume a well-mixed environment with mass action
kinetics. The fish are constantly swimming so this does not seem unreasonable.
We endow fish with a salmon louse count, $n$, which is initially zero, which we
indicate with a subscript, e.g. $F_n$. The first attachment depends only on the
fish \resizebox{!}{\baselineskip}{\salmon} and the presence of a nearby sea
louse \raisebox{0.25\baselineskip}{\resizebox{!}{0.6\baselineskip}{\copepod}},
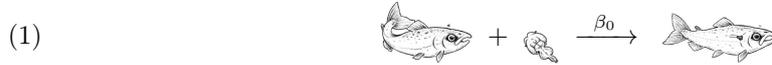
\begin{equation}
  \salmon{} + \copepod{} \xrightarrow{\;\;\beta_0\;\;}\;\isalmon{}
  \label{eq:infect}
\end{equation}
This is a rule, and is read in a similar way to a chemical reaction. A fish with zero attached sea lice and a sea louse interact together in such a way as to produce a fish with one attached sea louse \resizebox{!}{0.8\baselineskip}{\isalmon}. We could denote the reagents with letters, perhaps $F$ for fish and $C$ for sea lice (copepods) but instead we use drawings for clarity and aesthetic reasons. The rate at which this interaction happens, $\beta_{0}$, is unknown.

This rule is transformed into differential equations just as chemical reactions would be,
\begin{align}
  \frac{dF_{0}(t)}{dt} &= -\beta_{0}\left|C(t)\right|\left|F_{0}(t)\right| \\
  \frac{dF_{1}(t)}{dt} &= \beta_{0}\left|C(t)\right|\left|F_{0}(t)\right|
\end{align}
where the vertical bars, $\left| \cdot \right|$ give the count of the fish or sea lice as applicable. For the differential equation form,  we use more traditional notation.

In general, a fish with $n$ attached lice and a free louse can interact to produce a fish with $n+1$ attached lice, so we have, in fact, a family of rules parametrised by $n$.
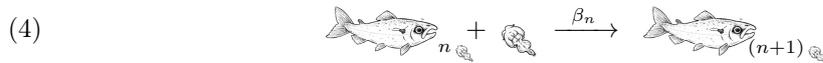
\begin{equation}
  \nsalmon{n}\hspace{-1em} + \copepod{} \xrightarrow{\;\;\beta_{n}\;\;}\;
  \nsalmon{\quad (n+1)}
  \label{eq:accumulate}
\end{equation}
It is perfectly possible --- and there is good reason to believe~\citep{ugelvik_atlantic_2017} --- that the attachment rate might differ according to the number of attached lice and for this reason we have a set of infection rates $\beta_{n}$ rather than a single one.

\subsection{Detachment of ectoparasites from fish}
As salmon lice can attach to fish, so can they detach (or experience mortality) at rate $\gamma$, and we can write this as the reverse of Rule~\ref{eq:accumulate},
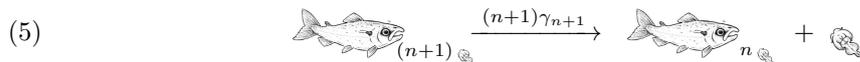
\begin{equation}
  \nsalmon{\quad(n+1)}
  \hspace{-1em}\xrightarrow{\;(n+1)\gamma_{n+1}\;}\;
  \nsalmon{n} + \copepod{}
  \label{eq:detach}
\end{equation}
again allowing that there might be different rates for different numbers of attached lice. The additional factor of $n+1$ arises because of a fact from combinatorics. There is only one way to add an element to a collection, we simply add it. However, to remove an element from a collection, we can remove any element that we choose. So \emph{any one} of the $n+1$ attached lice can detach (we consider these potential events to be mutually independent).

As written, a newly detached louse may re-attach to the same or another fish. Other dynamics are possible: it could also die, for example. Since we are using differential equations for simulation of attached counts on fish but the number of free lice is provided exogenously to this model by the dispersal models, we need not treat this question explicitly. In a more detailed stochastic model we would be required to be explicit about this choice.

\subsection{Free copepods}
We have not provided an account of copepods entering or leaving the cage or how long they remain within it. If we can assume that attachment and detachment (consuming and producing a free copepod respectively) have negligible effect on the ambient density, then we have no problem for the differential equation representation of the system: $C(t) = V\rho(t)$ where $V$ is the volume of the cage and $\rho(t)$ is provided by the hydrodynamic model. If we wished to do a stochastic simulation, where each copepod is explicitly counted, or if this assumption did not hold, we would require more information: the flow rate of the water through the cage, its surface area normal to the flow, its average diameter parallel to the flow.

\subsection{Numerical considerations}
Rules~\ref{eq:accumulate} and~\ref{eq:detach} form the entirety of the model. The differential equations for $F_{n}$ are,
\begin{align}
  \label{eq:odes}
  \frac{dF_{0}(t)}{dt} &= \gamma_{1}F_{1}(t) - \beta_{0}C(t)F_{0}(t) \nonumber\\
  \frac{dF_{n}(t)}{dt} &= \beta_{n-1}C(t)F_{n-1}(t) - \beta_{n}C(t)F_{n}(t) + (n+1)\gamma_{n+1}F_{n+1}(t)
\end{align}
as can be verified by inspection (and where the vertical bars have been dropped).
There is no \emph{a priori} limit on the number of attached lice so this system of differential equations has infinite dimensionality. This poses a practical problem for numerical integration: we need $n$ to be finite. We simply truncate the system at a suitably large value of $n$ --- double the largest count observed in the empirical data.

The $2n$ parameters, $\beta_{n}$ and $\gamma_{n}$, are not known and difficult to determine experimentally so they must be fit. If we consider each to be distinct, the number of parameters becomes far too large to practically estimate and, in any case, underdetermined by the relatively sparse data. We consider some simple cases below, when $\beta_{0} = \beta_{1} = \ldots = \beta$ and when $\beta_{1} = \beta_{2} = \ldots = \beta_{\text{acc}}$ but $\beta_{0} = \beta_{\text{ini}} \neq \beta_{\text{acc}}$. We additionally suppose a constant detachment rate, $\gamma_{1} = \gamma_{2} = \ldots = \gamma $.

\section{Results}

Having defined this dynamical system, let us see how it behaves. Figure~\ref{fig:distribution-time-shift} (top) illustrates the time-varying distribution produced for reasonable parameter values. It shows the shape of the distribution for finding a fish with exactly $n$ attached lice at intervals of 72 hours from the initial state. Here, the ambient copepod density is held constant. At first, zero fish are infected; we begin with a Dirac delta distribution at zero. As time progresses, the probability mass shifts towards higher values, it is more likely that a fish chosen at random has more attached lice. Figure~\ref{fig:distribution-time-shift} (bottom) shows a different view of the same distributions; it shows how the probabilities vary in time for different counts. Observe that the system eventually comes to an equilibrium and the distribution stabilises.
\begin{figure}
  \centering
  \begin{tikzpicture}
    \begin{axis}[
      cycle list name=exotic,
      width=0.8\textwidth, height=0.4\textwidth,
      ybar interval, minor y tick num = 3,
      xmin=0, xmax=10,
      ymin=0, ymax=1,
      title={Time-varying distribution for attached salmon louse count},
      xlabel={$n$ -- attached louse count},
      ylabel={Probability}
      ]
      \addplot table[x=n,y=t0] {data/simulated/density-sweep-b0.02-g0.01-d0.1.tsv};
      \addlegendentry{$t=0\text{ days }$}
      \addplot table[x=n,y=t72] {data/simulated/density-sweep-b0.02-g0.01-d0.1.tsv};
      \addlegendentry{$t=3\text{ days }$}
      \addplot table[x=n,y=t144] {data/simulated/density-sweep-b0.02-g0.01-d0.1.tsv};
      \addlegendentry{$t=6\text{ days }$}
      \addplot table[x=n,y=t216] {data/simulated/density-sweep-b0.02-g0.01-d0.1.tsv};
      \addlegendentry{$t=9\text{ days }$}
      \addplot table[x=n,y=t288] {data/simulated/density-sweep-b0.02-g0.01-d0.1.tsv};
      \addlegendentry{$t=12\text{ days }$}
      \addplot table[x=n,y=t360] {data/simulated/density-sweep-b0.02-g0.01-d0.1.tsv};
      \addlegendentry{$t=15\text{ days }$}
    \end{axis}
  \end{tikzpicture}
  \par
  \begin{tikzpicture}
    \begin{axis}[
      cycle list name=exotic,
      width=0.8\textwidth, height=0.4\textwidth,
      minor y tick num = 3,
      xmin=0, xmax=15,
      ymin=0, ymax=1,
      title={Time-varying probability for selected louse counts},
      xlabel={Days},
      ylabel={Probability}
      ]
      \addplot table[x expr=\thisrow{t}/24,y=f0] {data/simulated/density-timeseries.tsv};
      \addlegendentry{0 lice}
      \addplot table[x expr=\thisrow{t}/24,y=f1] {data/simulated/density-timeseries.tsv};
      \addlegendentry{1 louse}
      \addplot table[x expr=\thisrow{t}/24,y=f2] {data/simulated/density-timeseries.tsv};
      \addlegendentry{2 lice}
      \addplot table[x expr=\thisrow{t}/24,y=f3] {data/simulated/density-timeseries.tsv};
      \addlegendentry{3 lice}
      \addplot table[x expr=\thisrow{t}/24,y=f4] {data/simulated/density-timeseries.tsv};
      \addlegendentry{4 lice}
    \end{axis}
  \end{tikzpicture}
  \caption{\emph{Time-varying distributions.} (top) Distributions for finding up to 10 attached lice after 0, 3, $\ldots$, 15 days. Attachment rate is fixed, $\beta = 0.02$, detachment rate, $\gamma = 0.01$ and a density, $\rho(t) = 0.1$. (bottom) Probabilities over time for finding 0, 1, 2, 3 or 4 lice on a fish selected at random.}
  \label{fig:distribution-time-shift}
\end{figure}
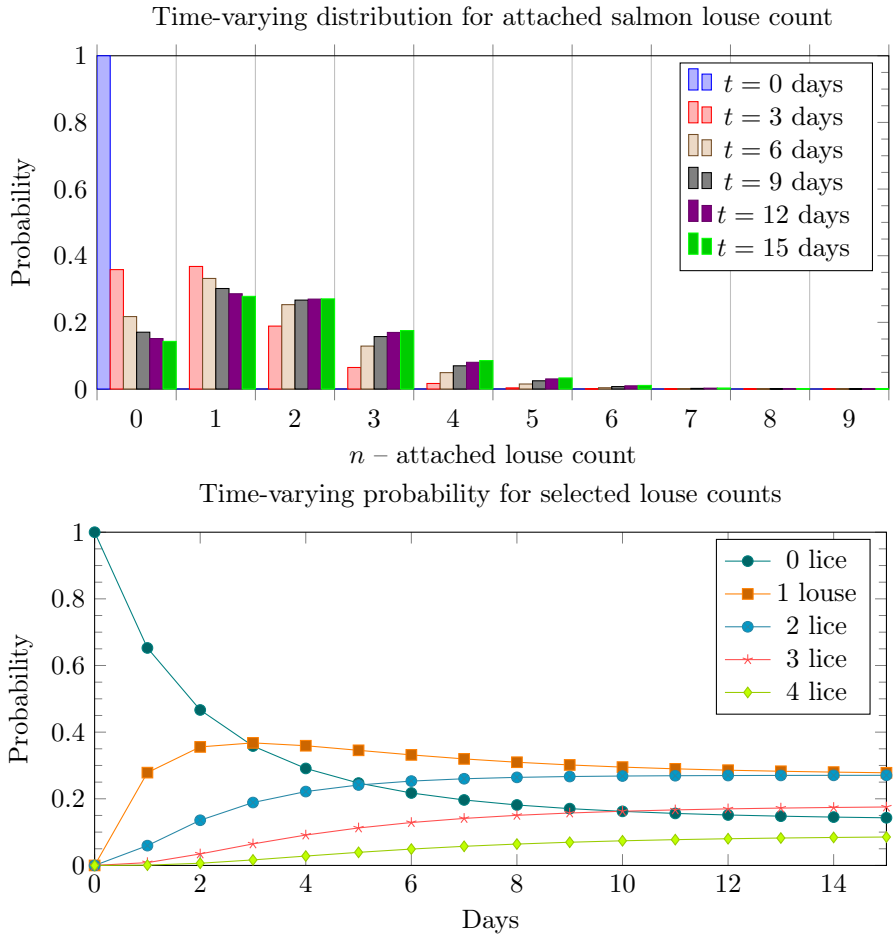

\begin{figure}
  \centering
  \begin{tikzpicture}
    \begin{axis}[
      cycle list name=exotic,
      width=0.8\textwidth, height=0.4\textwidth,
      ybar interval, minor y tick num = 3,
      xmin=0, xmax=10,
      ymin=0, ymax=1,
      title={Distribution for attached louse count after 15 days},
      xlabel={$n$ -- attached copepod count},
      ylabel={Probability}
      ]
      \addplot table[x=n,y=t360] {data/simulated/beta-sweep-b0.01-g0.01-d0.1.tsv};
      \addlegendentry{$\beta=0.01\text{ lice-fish}/\text{hr}$}
      \addplot table[x=n,y=t360] {data/simulated/beta-sweep-b0.02-g0.01-d0.1.tsv};
      \addlegendentry{$\beta=0.02\text{ lice-fish}/\text{hr}$}
      \addplot table[x=n,y=t360] {data/simulated/beta-sweep-b0.03-g0.01-d0.1.tsv};
      \addlegendentry{$\beta=0.03\text{ lice-fish}/\text{hr}$}
      \addplot table[x=n,y=t360] {data/simulated/beta-sweep-b0.04-g0.01-d0.1.tsv};
      \addlegendentry{$\beta=0.04\text{ lice-fish}/\text{hr}$}
      \addplot table[x=n,y=t360] {data/simulated/beta-sweep-b0.05-g0.01-d0.1.tsv};
      \addlegendentry{$\beta=0.05\text{ lice-fish}/\text{hr}$}
    \end{axis}
  \end{tikzpicture}
  \par
  \begin{tikzpicture}
    \begin{axis}[
      cycle list name=exotic,
      width=0.8\textwidth, height=0.4\textwidth,
      ybar interval, minor y tick num = 3,
      xmin=0, xmax=10,
      ymin=0, ymax=1,
      title={Distribution for attached louse count after 15 days},
      xlabel={$n$ -- attached louse count},
      ylabel={Probability}
      ]
      \addplot table[x=n,y=t360] {data/simulated/gamma-sweep-b0.02-g0.0-d0.1.tsv};
      \addlegendentry{$\gamma=0.0\text{ lice-fish}/\text{hr}$}
      \addplot table[x=n,y=t360] {data/simulated/gamma-sweep-b0.02-g0.005-d0.1.tsv};
      \addlegendentry{$\gamma=0.005\text{ lice-fish}/\text{hr}$}
      \addplot table[x=n,y=t360] {data/simulated/gamma-sweep-b0.02-g0.01-d0.1.tsv};
      \addlegendentry{$\gamma=0.01\text{ lice-fish}/\text{hr}$}
      \addplot table[x=n,y=t360] {data/simulated/gamma-sweep-b0.02-g0.015-d0.1.tsv};
      \addlegendentry{$\gamma=0.015\text{ lice-fish}/\text{hr}$}
      \addplot table[x=n,y=t360] {data/simulated/gamma-sweep-b0.02-g0.02-d0.1.tsv};
      \addlegendentry{$\gamma=0.02\text{ lice-fish}/\text{hr}$}
    \end{axis}
  \end{tikzpicture}
  \caption{\emph{Sensitivity to key parameters.} (top) Attachment rate $\beta$. Distributions after 15 days for different attachment rates. Higher attachment rates naturally lead, on average, to larger parasite burden; distributions are shifted increasingly to the right. Other parameters, detachment rate $\gamma=0.01$ and density $\rho(t)=0.1$ are held constant. (bottom) Detachment rate $\gamma$. As for the top panel, but with attachment rate held constant $\beta=0.02$. Larger values for $\gamma$ shift the distribution to the left as the lice are more likely to fall off.}
  \label{fig:sensitivity}
\end{figure}
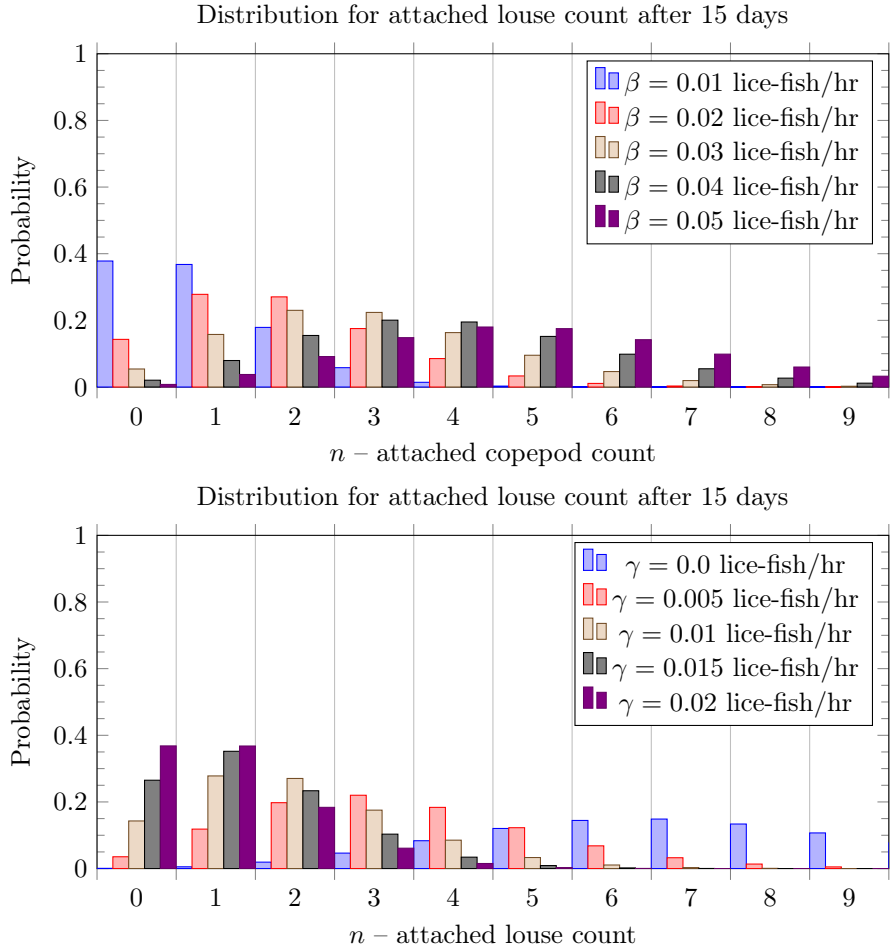

The two key parameters in the model are the attachment and detachment rates, $\beta$ and $\gamma$. We will fit these to empirical data below, but it is useful to understand, in general terms, the effects of these parameters on the model output. These parameters are characteristic of the two interacting organisms at a population level. In a real population these might vary with individual fish size or weight, the strength of their immune response, perhaps the temperature of the water and other factors. In Figure~\ref{fig:sensitivity} (top) we can see the effect of different values for the attachment rate, $\beta$. As might be expected, increased attachment rates result in greater burden on the fish who carry more lice. Increasing this parameter shifts the distributions (for $t = 15$ days in the figure) to the right, and flattens them somewhat. The detachment rate, $\gamma$ works in the opposite direction; increasing it shifts the distribution towards the left, giving the attached lice a greater propensity to fall off so when we pick a fish we are more likely to count fewer lice.

\subsection{Fitting against Loch Linnhe sentinel cages}

Fitting this model to real data is challenging. If we know the density of copepods in the water, we can proceed to fit values of $\beta$ and $\gamma$. Or, if we know the rate constants for the model, we can fit a time-averaged density. But we know none of these things and our ultimate enterprise is to validate densities that are derived from a hydrodynamic model. To extract ourselves from this quandry, we assert a constant reference density of 0.1 $\text{copepods}/m^{3}$ and ask what values the rate constants ought to take in this situation.

Fitting is done by comparing the empirical distributions of louse counts to the modelled distributions. Each cage deployment is considered independently: a probability distribution or normalised histogram is computed from the counts, the model is run starting with the appropriate number of uninfected fish for the duration of the deployment, and the resulting simulated distribution compared to the observed distribution. The comparison uses what is variously called the Earth Mover's Distance (EMD), the Vaserstein~\citep{vaserstein_markov_1969} or Kantorovich-Rubenstein~\citep{kantorovich_mathematical_1960} metric. The EMD provides a symmetric measure of the difference between two probability distributions (or any two functions whose integral over the domain of interest is equal). If two distributions are identical, the distance between them is zero, if they are not, the more different they are, the greater the distance. The EMD is therefore a suitable target for minimisation. We simply run the model with the number of fish found in the deployment for the duration of the deployment to obtain the expected distribution of louse counts, and compute the EMD between that distribution and the empirical one, and adjust parameter using a standard optimisation method.

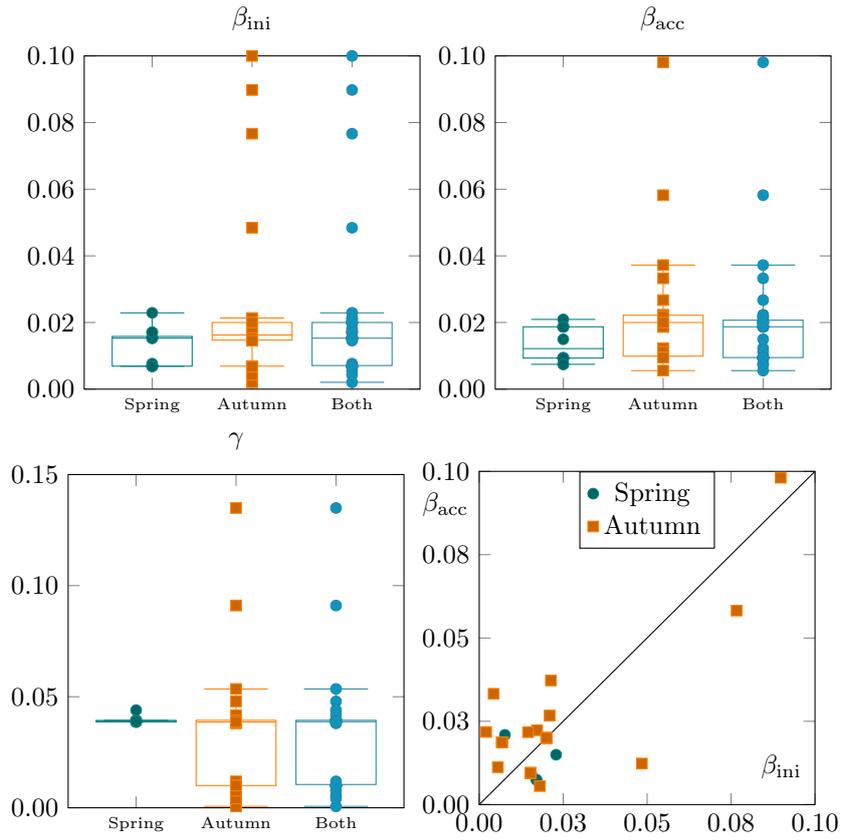
\begin{figure}
  \centering
  \begin{tikzpicture}
    \begin{axis}[
      cycle list name=exotic,
      boxplot,
      boxplot/draw direction=y,
      ymin=0,ymax=0.1,
      width=0.4\textwidth, height=0.4\textwidth,
      title={$\beta_{\text{ini}}$},
      xtick={1,2,3},
      xticklabels={Spring, Autumn, Both},
      xticklabel style={font=\tiny},
      scaled ticks=false,
      yticklabel style={
        /pgf/number format/precision=2,
        /pgf/number format/fixed,
        /pgf/number format/fixed zerofill,
      }
      ]
      \addplot+[only marks] table[x expr=1,y=beta0] {data/fitting/fitting-summary-spring.tsv};
      \pgfplotsset{cycle list shift=-1}
      \addplot+[no marks] table[y=beta0] {data/fitting/fitting-summary-spring.tsv};

      \addplot+[only marks] table[x expr=2,y=beta0] {data/fitting/fitting-summary-autumn.tsv};
      \pgfplotsset{cycle list shift=-2}
      \addplot+[no marks] table[y=beta0] {data/fitting/fitting-summary-autumn.tsv};

      \addplot+[only marks] table[x expr=3,y=beta0] {data/fitting/fitting-summary.tsv};
      \pgfplotsset{cycle list shift=-3}
      \addplot+[no marks] table[y=beta0] {data/fitting/fitting-summary.tsv};
    \end{axis}
  \end{tikzpicture}
  \begin{tikzpicture}
    \begin{axis}[
      cycle list name=exotic,
      boxplot,
      boxplot/draw direction=y,
      ymin=0,ymax=0.1,
      width=0.4\textwidth, height=0.4\textwidth,
      title={$\beta_{\text{acc}}$},
      xtick={1,2,3},
      xticklabels={Spring, Autumn, Both},
      xticklabel style={font=\tiny},
      scaled ticks=false,
      yticklabel style={
        /pgf/number format/precision=2,
        /pgf/number format/fixed,
        /pgf/number format/fixed zerofill,
      }
      ]
      \addplot+[only marks] table[x expr=1,y=beta1] {data/fitting/fitting-summary-spring.tsv};
      \pgfplotsset{cycle list shift=-1}
      \addplot+[no marks] table[y=beta1] {data/fitting/fitting-summary-spring.tsv};

      \addplot+[only marks] table[x expr=2,y=beta1] {data/fitting/fitting-summary-autumn.tsv};
      \pgfplotsset{cycle list shift=-2}
      \addplot+[no marks] table[y=beta1] {data/fitting/fitting-summary-autumn.tsv};

      \addplot+[only marks] table[x expr=3,y=beta1] {data/fitting/fitting-summary.tsv};
      \pgfplotsset{cycle list shift=-3}
      \addplot+[no marks] table[y=beta1] {data/fitting/fitting-summary.tsv};
    \end{axis}
  \end{tikzpicture}
  \par
  \begin{tikzpicture}
    \begin{axis}[
      cycle list name=exotic,
      boxplot,
      boxplot/draw direction=y,
      ymin=0, ymax=0.15,
      width=0.4\textwidth, height=0.4\textwidth,
      title={$\gamma$},
      xtick={1,2,3},
      xticklabels={Spring, Autumn, Both},
      xticklabel style={font=\tiny},
      scaled ticks=false,
      yticklabel style={
        /pgf/number format/precision=2,
        /pgf/number format/fixed,
        /pgf/number format/fixed zerofill,
      }
      ]
      \addplot+[only marks] table[x expr=1,y=gamma] {data/fitting/fitting-summary-spring.tsv};
      \pgfplotsset{cycle list shift=-1}
      \addplot+[no marks] table[y=gamma] {data/fitting/fitting-summary-spring.tsv};

      \addplot+[only marks] table[x expr=2,y=gamma] {data/fitting/fitting-summary-autumn.tsv};
      \pgfplotsset{cycle list shift=-2}
      \addplot+[no marks] table[y=gamma] {data/fitting/fitting-summary-autumn.tsv};

      \addplot+[only marks] table[x expr=3,y=gamma] {data/fitting/fitting-summary.tsv};
      \pgfplotsset{cycle list shift=-3}
      \addplot+[no marks] table[y=gamma] {data/fitting/fitting-summary.tsv};
    \end{axis}
  \end{tikzpicture}
  \begin{tikzpicture}
    \begin{axis}[
      width=0.4\textwidth, height=0.4\textwidth,
      cycle list name=exotic,
      xlabel={$\beta_{\text{ini}}$},
      ylabel={$\beta_{\text{acc}}$},
      xmin=0.0, ymin=0.0,
      xmax=0.1, ymax=0.1,
      xtick={0,0.025, 0.05, 0.075, 0.1},
      ytick={0,0.025, 0.05, 0.075, 0.1},
      scaled ticks=false,
      ticklabel style={
        /pgf/number format/precision=2,
        /pgf/number format/fixed,
        /pgf/number format/fixed zerofill,
      },
      every axis x label/.style=
            {at={(axis cs:0.09,0.005)}, anchor=south},
      every axis y label/.style=
            {at={(ticklabel cs:0.9,0.1)}, anchor=west},
      legend style={at={(axis cs:0.05,0.1)}, anchor=north},
      ]
      \addplot+[only marks] table[x=beta0,y=beta1] {data/fitting/fitting-summary-spring.tsv};
      \addlegendentry{Spring}
      \addplot+[only marks] table[x=beta0,y=beta1] {data/fitting/fitting-summary-autumn.tsv};
      \addlegendentry{Autumn}
      \draw (axis cs:0,0) -- (axis cs:0.1,0.1);
    \end{axis}
  \end{tikzpicture}
  \caption{Fit values for $\beta_{\text{ini}}$, $\beta_{\text{acc}}$, and $\gamma$ across the whole of the Loch Linnhe dataset. Infection burden is substantially different between spring and autumn deployments, so we show these separately as well as the combined values.}
  \label{fig:fit}
\end{figure}

The results of carrying out this fitting exercise against the Loch Linnhe sentinel cage data~\citep{pert_loch_2021} are shown in Figure~\ref{fig:fit}. An immediate observation is a much greater variance in the fit values in the autumn. In the spring, fewer fish are infected and those that are have a much lower infection burden than in the autumn. This variance is most likely due to our simplistic assumption of a constant density of free copepods. In the autumn data, we can observe the presence of the effect described by \citet{ugelvik_atlantic_2017}, where propensity to become infected increases when already infected. In the lower-right panel, the scatterplot of $\beta_{\text{ini}}$ vs $\beta_{\text{acc}}$, there is a tendency for points to lie above the line of $\beta_{\text{ini}} = \beta_{\text{acc}}$.

Finally, we inspect a few cages, the empirical distributions of louse counts and the simulated distributions with fit values. Plots for all cages are included in Appendix~\ref{sec:allcages}. Figure~\ref{fig:happyfits} shows two such cages with developed sea lice infestations in the autumn. These scenarios are representative of the situation where the model works well. Originally there would have been 50 fish in these cages, about 20\% of which likely died or were otherwise lost. Of the remaining fish, we can see that the simulated distributions of counts are qualitatively similar to the observed distributions. Note also the parameter values: the increased rate of infection for already infected fish is much higher than the initial rate of infection, and the rate of lice detaching is much lower.
\begin{figure}
  \begin{tikzpicture}
  \begin{axis}[
    cycle list name=exotic,
    width=0.48\textwidth, height=0.4\textwidth,
    ybar interval, minor y tick num = 3,
    xmin=-0.5, xmax=11.500000,
    ymin=0, ymax=1,
    xminorgrids=true,
    xtick={0,5,...,40},
    minor x tick num=4,
    x tick label as interval=false,
    bar width=0.4,
    title={Cage 1 autumn 2013 (13 days)},
    xlabel={$n$ -- attached copepod count},
    ylabel={Probability}
    ]
    \addplot +[ybar,bar shift=-0.2] table[x=n,y=p] {data/simulated/sim_1_2013-10-15.tsv};
    \addlegendentry{Observed}
    \addplot +[ybar,bar shift=0.2] table[x=n,y=q] {data/simulated/sim_1_2013-10-15.tsv};
    \addlegendentry{Simulated}
    \node at (axis cs:8.800000,0.5) {\begin{minipage}[t]{2cm}
      \small
      \begin{align*}
        \beta_\text{ini} &= 0.0056\\
        \beta_\text{acc} &= 0.0112\\
        \gamma &= 0.0006\\
        41 & \text{ fish}
      \end{align*}
    \end{minipage}};
  \end{axis}
\end{tikzpicture}
  \begin{tikzpicture}
  \begin{axis}[
    cycle list name=exotic,
    width=0.48\textwidth, height=0.4\textwidth,
    ybar interval, minor y tick num = 3,
    xmin=-0.5, xmax=16.500000,
    ymin=0, ymax=1,
    xminorgrids=true,
    xtick={0,5,...,40},
    minor x tick num=4,
    x tick label as interval=false,
    bar width=0.4,
    title={Cage 6 autumn 2011 (14 days)},
    xlabel={$n$ -- attached copepod count},
    ylabel={Probability}
    ]
    \addplot +[ybar,bar shift=-0.2] table[x=n,y=p] {data/simulated/sim_6_2011-11-09.tsv};
    \addlegendentry{Observed}
    \addplot +[ybar,bar shift=0.2] table[x=n,y=q] {data/simulated/sim_6_2011-11-09.tsv};
    \addlegendentry{Simulated}
    \node at (axis cs:12.800000,0.5) {\begin{minipage}[t]{2cm}
      \small
      \begin{align*}
        \beta_\text{ini} &= 0.0043\\
        \beta_\text{acc} &= 0.0333\\
        \gamma &= 0.0054\\
        42 & \text{ fish}
      \end{align*}
    \end{minipage}};
  \end{axis}
\end{tikzpicture}
  \caption{Observed and simulated data, annotated with fit parameter values for two cage deployments; cages 1 and 6, both in the autumn.}
  \label{fig:happyfits}
\end{figure}
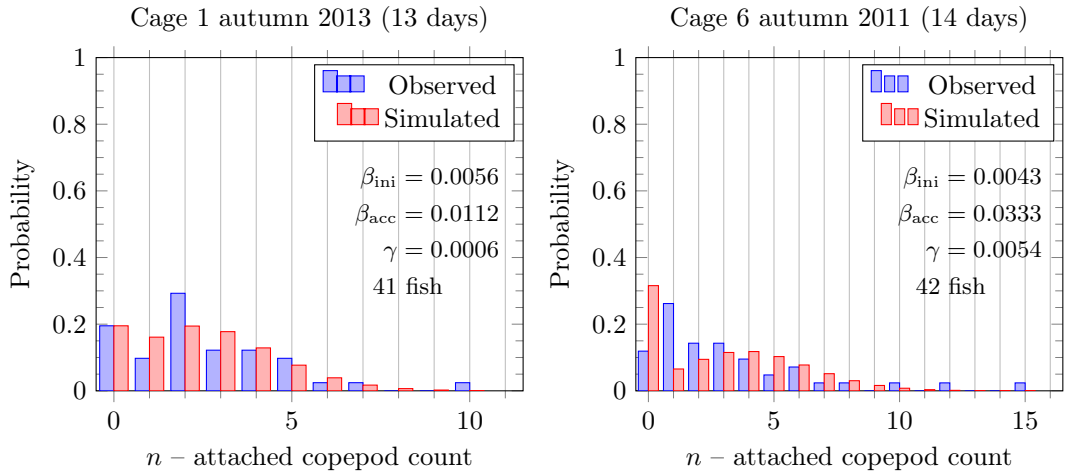

The particular numerical values for these quantities in Figure~\ref{fig:happyfits} are quite different from one another. This is explained by our assumption of a constant density ($\rho(t) = 0.1\text{ copepods }/m^{3}$) in all circumstances which surely was not the case during these deployments. The substantially higher infection burden experienced by cage 6 in the autumn of 2011 suggests a higher density. Since the attachment propensities are linear in this density, we might conclude that the true density in this case was, on average, about 2-3$\times$ higher than that experienced by cage 1 in the autumn of 2013.

\begin{figure}
  \begin{tikzpicture}
  \begin{axis}[
    cycle list name=exotic,
    width=0.48\textwidth, height=0.4\textwidth,
    ybar interval, minor y tick num = 3,
    xmin=-0.5, xmax=10.500000,
    ymin=0, ymax=1,
    xminorgrids=true,
    xtick={0,5,...,40},
    minor x tick num=4,
    x tick label as interval=false,
    bar width=0.4,
    title={Cage 2 spring 2011 (15 days)},
    xlabel={$n$ -- attached copepod count},
    ylabel={Probability}
    ]
    \addplot +[ybar,bar shift=-0.2] table[x=n,y=p] {data/simulated/sim_2_2011-05-27.tsv};
    \addlegendentry{Observed}
    \addplot +[ybar,bar shift=0.2] table[x=n,y=q] {data/simulated/sim_2_2011-05-27.tsv};
    \addlegendentry{Simulated}
    \node at (axis cs:8.000000,0.5) {\begin{minipage}[t]{2cm}
      \small
      \begin{align*}
        \beta_\text{ini} &= 0.0153\\
        \beta_\text{acc} &= 0.0093\\
        \gamma &= 0.0393\\
        39 & \text{ fish}
      \end{align*}
    \end{minipage}};
  \end{axis}
\end{tikzpicture}
  \begin{tikzpicture}
  \begin{axis}[
    cycle list name=exotic,
    width=0.48\textwidth, height=0.4\textwidth,
    ybar interval, minor y tick num = 3,
    xmin=-0.5, xmax=37.500000,
    ymin=0, ymax=1,
    xminorgrids=true,
    xtick={0,5,...,40},
    minor x tick num=4,
    x tick label as interval=false,
    bar width=0.4,
    title={Cage 3B autumn 2011 (14 days)},
    xlabel={$n$ -- attached copepod count},
    ylabel={Probability}
    ]
    \addplot +[ybar,bar shift=-0.2] table[x=n,y=p] {data/simulated/sim_3B_2011-11-09.tsv};
    \addlegendentry{Observed}
    \addplot +[ybar,bar shift=0.2] table[x=n,y=q] {data/simulated/sim_3B_2011-11-09.tsv};
    \addlegendentry{Simulated}
    \node at (axis cs:29.600000,0.5) {\begin{minipage}[t]{2cm}
      \small
      \begin{align*}
        \beta_\text{ini} &= 0.1000\\
        \beta_\text{acc} &= 0.3000\\
        \gamma &= 0.0100\\
        34 & \text{ fish}
      \end{align*}
    \end{minipage}};
  \end{axis}
\end{tikzpicture}
  \caption{Problematic cases for parameter fitting.}
  \label{fig:sadfits}
\end{figure}
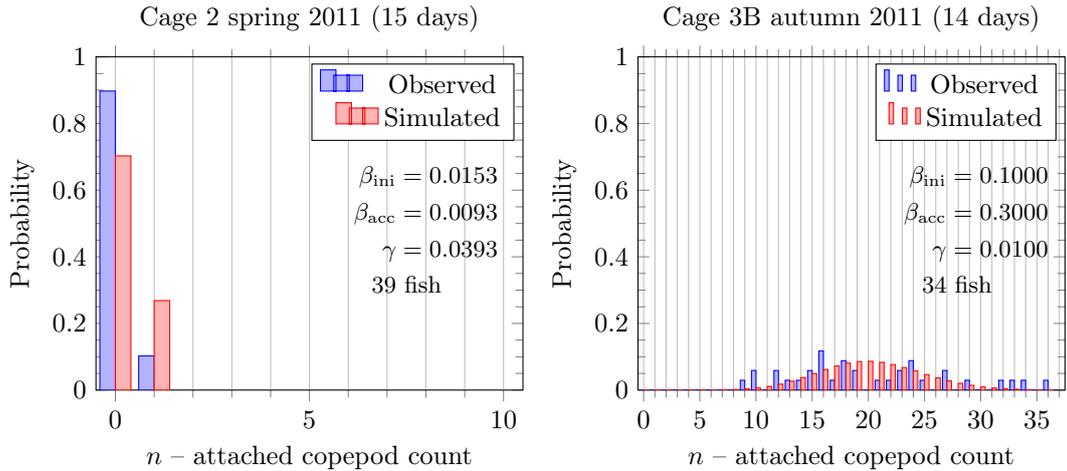
Figure~\ref{fig:sadfits} shows two representative cases where the fitting operation fails. On the left, there is very little signal in the data. Fully 90\% of fish are uninfected and the remaining 10\% have only a single attached parasite. There is no evidence of accumulation in the data so clearly any attempt to estimate $\beta_{\text{acc}}$ cannot work. The other parameters are also underconstrained by the data. It would be possible to compute the entropy of the empirical distribution to distinguish between data that can sensibly be used for fitting and data that has insufficient statistical power. This situation appears to be typical of spring deployments of sentinel cages in this dataset.

The right-hand panel shows a complete failure of the optimisation process under conditions of very high infection levels. The values for $\beta_{\text{ini}} = 0.1$ and $\beta_{\text{acc}} = 0.3$ were obtained by hand. Whilst it is possible to drive this model to a regime where it produces a distribution in the right neighbourhood, the values are an order of magnitude greater than what we find for more ``typical'' deployments. A similar effect can be observed with cages 5, 8, and 10 that same year (see Appendix~\ref{sec:allcages}). This suggests significantly different conditions than the assumed baseline density of 0.1 sea lice per m$^2$.

\section{Discussion}

In this paper, we have shown a way to connect copepod density estimates such as are produced by hydrodynamic models to the infection burden expressed as expected distributions of louse count on salmon in cages using a semi-mechanistic model with, effectively, two or three parameters. Starting from an assumption of constant density, we can estimate the parameter values which, given that density, accurately reproduce the empirical distributions of infection burden. In many instances of the data from Loch Linnhe~\citep{pert_loch_2021}, this strategy is remarkably effective. The project of validating hydrodynamic models can now be cast as: does using time-varying density data from a hydrodynamic model allow for a better fit than using a constant density? Future work will address this question.

The model we present here was produced as part of a larger project to construct a software pipeline to automate this process. The pipeline allows modellers to evaluate time-varying copepod densities against observed counts, and adjust models in response to pipeline output as well as comparison between models via the fit (interpreted as ``implied'') rate constants. To support reproducibility and transparency, the latter especially important if technology such as this is used in a regulatory context, the pipeline records provenance metadata throughout.

Beyond simple validation, given reliable estimates of the numerical values for the rate constants (the benchmark set by this paper puts $\beta_{\text{ini}} \approx 0.018$, $\beta_{\text{acc}} \approx 0.036$, and $\gamma \approx 0.002$ but these can surely be improved) and confidence in the accuracy of hydrodynamic models, the model given here can be used predictively. At any point in the density field, place an instance of this model and it will say, at any time, what distribution of infection burden we ought to oberve were a sentinel cage placed in that spot. There are, of course, limits. We have no information from the data used here about the natural history of infection beyond about two weeks, so that is a reasonable time horizon beyond which prediction accuracy would be questionable. However, this capability illustrates the strength of the semi-mechanistic approach compared to a statistical approach of just fitting a zero-inflated distribution to the data without an account, however simplified, of its biological genesis or how it should be expected to change in time.


The system as described here assumes that individual fish (and, indeed, salmon lice) have identical characteristics. In reality, individuals differ in potentially relevant ways. Larger fish can acquire a greater infection burden by virtue of their greater surface area to which parasites might attach. To address this, the model could be altered to work with a stratified population. This alteration is done in the obvious way, via a construction called a ``pullback''~\citep{libkind_algebraic_2022,baez_compositional_2023}, just as a stratified SIR epidemic model articulated as a bilinear map is obtained from the standard homogeneous formulation. However, doing so results in a proliferation of parameters. Rather than having $\beta_{0}, \beta_{1}, \ldots$, we have the matrix $\beta_{nm}$ where $n$ is the louse count as before, and $m$ is an index for the stratum or size-band. Stratifying according to other characteristics such as where on the fish parasites attach is likewise possible at a similar cost.

Though this model is sufficiently simple to directly derive the differential equations (\ref{eq:odes}), it is not difficult to imagine a more elaborate model where this would not be practical. Incorporating a detailed account of immune response of the fish and its effect of shedding parasites is possible in principle and might be interesting. Equally, distinguishing parasites by developmental stage and including different behaviours in different stages is likely to be fruitful. We do not pass judgement on what data would be required for sufficient statistical power to calibrate such a model or the computational expense of doing so and merely point out that it is straightforward to \emph{express} more complex models in the framework we use here.  The way that a master equation is encoded in the rules~\ref{eq:accumulate}-\ref{eq:detach} uses a relatively new feature of the Kappa modelling language~\citep{boutillier_kappa_2018} called ``counters''~\citep{boutillier_counters_2019}.  Our earlier work on COVID-19~\citep{waites_compositional_2022} also used this feature but not as systematically as we have done here. A reader with knowledge of statistical physics or quantum mechanics might remark on the presence of the factor of $n$ in the rate for rule~\ref{eq:detach} and identify it with the same factor that appears on application of an annihilation operator~\citep{baez_quantum_2018}. The capability of encoding a master equation in rules using counters as we have done here remains available even when the model is complex enough that it is no longer convenient to use the differential equation form. It simply becomes necessary to simulate the system stochastically to approximate the trajectories of probability distributions, at greater computational expense.

A final remark on this formulation in terms of a trajectory of distributions is in order. The distribution trajectory produced by the differential equations corresponds to the large population limit, in a sense, the \emph{average} distribution at each point in time. With a small population such as we have here, at most 50 fish and copepod densities in the water on the order of $0.1/m^{3}$, it is unreasonable to expect that sampling of any real cage deployment should produce exactly the average distribution. If we do not use the convenient shortcut of deriving the differential equations and instead simulate the system stochastically, we can then sample trajectories of distributions and obtain a distribution of trajectories of distributions. This is a well-defined concept since we know how to quantify the distance between distributions and even determine how likely it is to find a particular distribution in the data a certain distance from the average. We do not do so here because of the limited sentinel cage data available. Nevertheless, this is an interesting avenue for future investigation.

\section{Code and data availability}

Source code underpinning this paper is available at \url{https://codeberg.org/rbem/copepod-master}. It is written in the Python language in order to be as broadly accessible as possible. For those interested in using this technique in some other language, the best place to start is most likely the \texttt{dFdt()} function in \texttt{src/cmode.py}. It is simple and can easily be translated. This is used by \texttt{src/cmsim.py} which integrates the differential equation system with provided parameters (including the ability to use a data file containing densities for exogenous forcing, not used in this paper), as well as \texttt{src/cmfit.py} which does parameter estimation. The remainder of the code is about pre- and post-processing data, \texttt{src/pertsplit.py} splits the source data into separate files for each cage deployment, \texttt{src/linnhe.py} produces probability distributions from counts and provides a couple of utility functions for accessing data about a particular deployment. The program \texttt{src/aplot.py} produces the plots of Figures~\ref{fig:happyfits} and~\ref{fig:sadfits} as well as Appendix~\ref{sec:allcages}. The only exotic dependency is on the \href{https://scipy.org/}{SciPy} library for scientific computing. Processing the data to produce this publication is driven by a \texttt{Makefile} in the top-level directory.

A version of the model written in the KaSim dialect of the Kappa language~\citep{boutillier_kappa_2018} is provided in \texttt{src/cm.ska}. This file is pre-processed with the \texttt{src/stratify.py} utility to produce \texttt{src/cm.ka} which can be run with the KaSim simulation tool. Of necessity, this version of the model has extra rules pertaining to the appearance of copepods at the boundary of a cage and their traversal within it, as well as an extra ``diameter'' parameter to complete the geometry of the cage. This version of the model provides for stochastic simulation of the system and straightforward extensibility.

The data used for fitting is that of~\citet{pert_loch_2021} available from \url{https://data.marine.gov.scot/dataset/loch-linnhe-biological-sampling-data-products-2011-2013-0}. A verbatim copy accompanies the source code for convenience in the file\\
\texttt{data/Sentinel\_cage\_sampling\_info\_update\_01122022.csv}. Processing to produce the data-driven results section of this paper begins from that file.

\section{Acknowledgements}

WW and DG were supported by a grant from the Sustainable Aquaculture Innovation Centre, Stirling. WW received funding from MRC grant MR/X011658/1 (\emph{Rule-based epidemic modelling}).

\setlength{\bibsep}{0pt}
\bibliographystyle{abbrvnat}
\bibliography{refs}

\clearpage\appendix
\section{Fitting results for all sentinel cages}
\label{sec:allcages}
\subsection{Cage 1}
\begin{center}
  \begin{tikzpicture}
  \begin{axis}[
    cycle list name=exotic,
    width=0.48\textwidth, height=0.4\textwidth,
    ybar interval, minor y tick num = 3,
    xmin=-0.5, xmax=10.500000,
    ymin=0, ymax=1,
    xminorgrids=true,
    xtick={0,5,...,40},
    minor x tick num=4,
    x tick label as interval=false,
    bar width=0.4,
    title={Cage 1 autumn 2011 (14 days)},
    xlabel={$n$ -- attached copepod count},
    ylabel={Probability}
    ]
    \addplot +[ybar,bar shift=-0.2] table[x=n,y=p] {data/simulated/sim_1_2011-11-09.tsv};
    \addlegendentry{Observed}
    \addplot +[ybar,bar shift=0.2] table[x=n,y=q] {data/simulated/sim_1_2011-11-09.tsv};
    \addlegendentry{Simulated}
    \node at (axis cs:8.000000,0.5) {\begin{minipage}[t]{2cm}
      \small
      \begin{align*}
        \beta_\text{ini} &= 0.0021\\
        \beta_\text{acc} &= 0.0218\\
        \gamma &= 0.1350\\
        1 & \text{ fish}
      \end{align*}
    \end{minipage}};
  \end{axis}
\end{tikzpicture}
  \begin{tikzpicture}
  \begin{axis}[
    cycle list name=exotic,
    width=0.48\textwidth, height=0.4\textwidth,
    ybar interval, minor y tick num = 3,
    xmin=-0.5, xmax=10.500000,
    ymin=0, ymax=1,
    xminorgrids=true,
    xtick={0,5,...,40},
    minor x tick num=4,
    x tick label as interval=false,
    bar width=0.4,
    title={Cage 1 spring 2012 (16 days)},
    xlabel={$n$ -- attached copepod count},
    ylabel={Probability}
    ]
    \addplot +[ybar,bar shift=-0.2] table[x=n,y=p] {data/simulated/sim_1_2012-05-17.tsv};
    \addlegendentry{Observed}
    \addplot +[ybar,bar shift=0.2] table[x=n,y=q] {data/simulated/sim_1_2012-05-17.tsv};
    \addlegendentry{Simulated}
    \node at (axis cs:8.000000,0.5) {\begin{minipage}[t]{2cm}
      \small
      \begin{align*}
        \beta_\text{ini} &= 0.0069\\
        \beta_\text{acc} &= 0.0187\\
        \gamma &= 0.0387\\
        30 & \text{ fish}
      \end{align*}
    \end{minipage}};
  \end{axis}
\end{tikzpicture}\\
  \begin{tikzpicture}
  \begin{axis}[
    cycle list name=exotic,
    width=0.48\textwidth, height=0.4\textwidth,
    ybar interval, minor y tick num = 3,
    xmin=-0.5, xmax=10.500000,
    ymin=0, ymax=1,
    xminorgrids=true,
    xtick={0,5,...,40},
    minor x tick num=4,
    x tick label as interval=false,
    bar width=0.4,
    title={Cage 1 autumn 2012 (16 days)},
    xlabel={$n$ -- attached copepod count},
    ylabel={Probability}
    ]
    \addplot +[ybar,bar shift=-0.2] table[x=n,y=p] {data/simulated/sim_1_2012-10-19.tsv};
    \addlegendentry{Observed}
    \addplot +[ybar,bar shift=0.2] table[x=n,y=q] {data/simulated/sim_1_2012-10-19.tsv};
    \addlegendentry{Simulated}
    \node at (axis cs:8.000000,0.5) {\begin{minipage}[t]{2cm}
      \small
      \begin{align*}
        \beta_\text{ini} &= 0.0153\\
        \beta_\text{acc} &= 0.0095\\
        \gamma &= 0.0394\\
        50 & \text{ fish}
      \end{align*}
    \end{minipage}};
  \end{axis}
\end{tikzpicture}
  \begin{tikzpicture}
  \begin{axis}[
    cycle list name=exotic,
    width=0.48\textwidth, height=0.4\textwidth,
    ybar interval, minor y tick num = 3,
    xmin=-0.5, xmax=11.500000,
    ymin=0, ymax=1,
    xminorgrids=true,
    xtick={0,5,...,40},
    minor x tick num=4,
    x tick label as interval=false,
    bar width=0.4,
    title={Cage 1 autumn 2013 (13 days)},
    xlabel={$n$ -- attached copepod count},
    ylabel={Probability}
    ]
    \addplot +[ybar,bar shift=-0.2] table[x=n,y=p] {data/simulated/sim_1_2013-10-15.tsv};
    \addlegendentry{Observed}
    \addplot +[ybar,bar shift=0.2] table[x=n,y=q] {data/simulated/sim_1_2013-10-15.tsv};
    \addlegendentry{Simulated}
    \node at (axis cs:8.800000,0.5) {\begin{minipage}[t]{2cm}
      \small
      \begin{align*}
        \beta_\text{ini} &= 0.0056\\
        \beta_\text{acc} &= 0.0112\\
        \gamma &= 0.0006\\
        41 & \text{ fish}
      \end{align*}
    \end{minipage}};
  \end{axis}
\end{tikzpicture}
\end{center}
\subsection{Cage 2}
\begin{center}
  \begin{tikzpicture}
  \begin{axis}[
    cycle list name=exotic,
    width=0.48\textwidth, height=0.4\textwidth,
    ybar interval, minor y tick num = 3,
    xmin=-0.5, xmax=10.500000,
    ymin=0, ymax=1,
    xminorgrids=true,
    xtick={0,5,...,40},
    minor x tick num=4,
    x tick label as interval=false,
    bar width=0.4,
    title={Cage 2 spring 2011 (15 days)},
    xlabel={$n$ -- attached copepod count},
    ylabel={Probability}
    ]
    \addplot +[ybar,bar shift=-0.2] table[x=n,y=p] {data/simulated/sim_2_2011-05-27.tsv};
    \addlegendentry{Observed}
    \addplot +[ybar,bar shift=0.2] table[x=n,y=q] {data/simulated/sim_2_2011-05-27.tsv};
    \addlegendentry{Simulated}
    \node at (axis cs:8.000000,0.5) {\begin{minipage}[t]{2cm}
      \small
      \begin{align*}
        \beta_\text{ini} &= 0.0153\\
        \beta_\text{acc} &= 0.0093\\
        \gamma &= 0.0393\\
        39 & \text{ fish}
      \end{align*}
    \end{minipage}};
  \end{axis}
\end{tikzpicture}
  \begin{tikzpicture}
  \begin{axis}[
    cycle list name=exotic,
    width=0.48\textwidth, height=0.4\textwidth,
    ybar interval, minor y tick num = 3,
    xmin=-0.5, xmax=10.500000,
    ymin=0, ymax=1,
    xminorgrids=true,
    xtick={0,5,...,40},
    minor x tick num=4,
    x tick label as interval=false,
    bar width=0.4,
    title={Cage 2 autumn 2011 (14 days)},
    xlabel={$n$ -- attached copepod count},
    ylabel={Probability}
    ]
    \addplot +[ybar,bar shift=-0.2] table[x=n,y=p] {data/simulated/sim_2_2011-11-09.tsv};
    \addlegendentry{Observed}
    \addplot +[ybar,bar shift=0.2] table[x=n,y=q] {data/simulated/sim_2_2011-11-09.tsv};
    \addlegendentry{Simulated}
    \node at (axis cs:8.000000,0.5) {\begin{minipage}[t]{2cm}
      \small
      \begin{align*}
        \beta_\text{ini} &= 0.0172\\
        \beta_\text{acc} &= 0.0223\\
        \gamma &= 0.0119\\
        50 & \text{ fish}
      \end{align*}
    \end{minipage}};
  \end{axis}
\end{tikzpicture}\\
  \begin{tikzpicture}
  \begin{axis}[
    cycle list name=exotic,
    width=0.48\textwidth, height=0.4\textwidth,
    ybar interval, minor y tick num = 3,
    xmin=-0.5, xmax=10.500000,
    ymin=0, ymax=1,
    xminorgrids=true,
    xtick={0,5,...,40},
    minor x tick num=4,
    x tick label as interval=false,
    bar width=0.4,
    title={Cage 2 spring 2012 (16 days)},
    xlabel={$n$ -- attached copepod count},
    ylabel={Probability}
    ]
    \addplot +[ybar,bar shift=-0.2] table[x=n,y=p] {data/simulated/sim_2_2012-05-17.tsv};
    \addlegendentry{Observed}
    \addplot +[ybar,bar shift=0.2] table[x=n,y=q] {data/simulated/sim_2_2012-05-17.tsv};
    \addlegendentry{Simulated}
    \node at (axis cs:8.000000,0.5) {\begin{minipage}[t]{2cm}
      \small
      \begin{align*}
        \beta_\text{ini} &= 0.0154\\
        \beta_\text{acc} &= 0.0094\\
        \gamma &= 0.0394\\
        33 & \text{ fish}
      \end{align*}
    \end{minipage}};
  \end{axis}
\end{tikzpicture}
  \begin{tikzpicture}
  \begin{axis}[
    cycle list name=exotic,
    width=0.48\textwidth, height=0.4\textwidth,
    ybar interval, minor y tick num = 3,
    xmin=-0.5, xmax=10.500000,
    ymin=0, ymax=1,
    xminorgrids=true,
    xtick={0,5,...,40},
    minor x tick num=4,
    x tick label as interval=false,
    bar width=0.4,
    title={Cage 2 autumn 2012 (16 days)},
    xlabel={$n$ -- attached copepod count},
    ylabel={Probability}
    ]
    \addplot +[ybar,bar shift=-0.2] table[x=n,y=p] {data/simulated/sim_2_2012-10-19.tsv};
    \addlegendentry{Observed}
    \addplot +[ybar,bar shift=0.2] table[x=n,y=q] {data/simulated/sim_2_2012-10-19.tsv};
    \addlegendentry{Simulated}
    \node at (axis cs:8.000000,0.5) {\begin{minipage}[t]{2cm}
      \small
      \begin{align*}
        \beta_\text{ini} &= 0.0153\\
        \beta_\text{acc} &= 0.0095\\
        \gamma &= 0.0394\\
        50 & \text{ fish}
      \end{align*}
    \end{minipage}};
  \end{axis}
\end{tikzpicture}\\
  \begin{tikzpicture}
  \begin{axis}[
    cycle list name=exotic,
    width=0.48\textwidth, height=0.4\textwidth,
    ybar interval, minor y tick num = 3,
    xmin=-0.5, xmax=11.500000,
    ymin=0, ymax=1,
    xminorgrids=true,
    xtick={0,5,...,40},
    minor x tick num=4,
    x tick label as interval=false,
    bar width=0.4,
    title={Cage 2 autumn 2013 (13 days)},
    xlabel={$n$ -- attached copepod count},
    ylabel={Probability}
    ]
    \addplot +[ybar,bar shift=-0.2] table[x=n,y=p] {data/simulated/sim_2_2013-10-15.tsv};
    \addlegendentry{Observed}
    \addplot +[ybar,bar shift=0.2] table[x=n,y=q] {data/simulated/sim_2_2013-10-15.tsv};
    \addlegendentry{Simulated}
    \node at (axis cs:8.800000,0.5) {\begin{minipage}[t]{2cm}
      \small
      \begin{align*}
        \beta_\text{ini} &= 0.0210\\
        \beta_\text{acc} &= 0.0267\\
        \gamma &= 0.0042\\
        46 & \text{ fish}
      \end{align*}
    \end{minipage}};
  \end{axis}
\end{tikzpicture}
\end{center}
\subsection{Cage 3B}
\begin{center}
  \begin{tikzpicture}
  \begin{axis}[
    cycle list name=exotic,
    width=0.48\textwidth, height=0.4\textwidth,
    ybar interval, minor y tick num = 3,
    xmin=-0.5, xmax=10.500000,
    ymin=0, ymax=1,
    xminorgrids=true,
    xtick={0,5,...,40},
    minor x tick num=4,
    x tick label as interval=false,
    bar width=0.4,
    title={Cage 3B spring 2011 (15 days)},
    xlabel={$n$ -- attached copepod count},
    ylabel={Probability}
    ]
    \addplot +[ybar,bar shift=-0.2] table[x=n,y=p] {data/simulated/sim_3B_2011-05-27.tsv};
    \addlegendentry{Observed}
    \addplot +[ybar,bar shift=0.2] table[x=n,y=q] {data/simulated/sim_3B_2011-05-27.tsv};
    \addlegendentry{Simulated}
    \node at (axis cs:8.000000,0.5) {\begin{minipage}[t]{2cm}
      \small
      \begin{align*}
        \beta_\text{ini} &= 0.0153\\
        \beta_\text{acc} &= 0.0093\\
        \gamma &= 0.0393\\
        48 & \text{ fish}
      \end{align*}
    \end{minipage}};
  \end{axis}
\end{tikzpicture}
  \begin{tikzpicture}
  \begin{axis}[
    cycle list name=exotic,
    width=0.48\textwidth, height=0.4\textwidth,
    ybar interval, minor y tick num = 3,
    xmin=-0.5, xmax=37.500000,
    ymin=0, ymax=1,
    xminorgrids=true,
    xtick={0,5,...,40},
    minor x tick num=4,
    x tick label as interval=false,
    bar width=0.4,
    title={Cage 3B autumn 2011 (14 days)},
    xlabel={$n$ -- attached copepod count},
    ylabel={Probability}
    ]
    \addplot +[ybar,bar shift=-0.2] table[x=n,y=p] {data/simulated/sim_3B_2011-11-09.tsv};
    \addlegendentry{Observed}
    \addplot +[ybar,bar shift=0.2] table[x=n,y=q] {data/simulated/sim_3B_2011-11-09.tsv};
    \addlegendentry{Simulated}
    \node at (axis cs:29.600000,0.5) {\begin{minipage}[t]{2cm}
      \small
      \begin{align*}
        \beta_\text{ini} &= 0.1000\\
        \beta_\text{acc} &= 0.3000\\
        \gamma &= 0.0100\\
        34 & \text{ fish}
      \end{align*}
    \end{minipage}};
  \end{axis}
\end{tikzpicture}\\
  \begin{tikzpicture}
  \begin{axis}[
    cycle list name=exotic,
    width=0.48\textwidth, height=0.4\textwidth,
    ybar interval, minor y tick num = 3,
    xmin=-0.5, xmax=10.500000,
    ymin=0, ymax=1,
    xminorgrids=true,
    xtick={0,5,...,40},
    minor x tick num=4,
    x tick label as interval=false,
    bar width=0.4,
    title={Cage 3B spring 2012 (16 days)},
    xlabel={$n$ -- attached copepod count},
    ylabel={Probability}
    ]
    \addplot +[ybar,bar shift=-0.2] table[x=n,y=p] {data/simulated/sim_3B_2012-05-17.tsv};
    \addlegendentry{Observed}
    \addplot +[ybar,bar shift=0.2] table[x=n,y=q] {data/simulated/sim_3B_2012-05-17.tsv};
    \addlegendentry{Simulated}
    \node at (axis cs:8.000000,0.5) {\begin{minipage}[t]{2cm}
      \small
      \begin{align*}
        \beta_\text{ini} &= 0.0077\\
        \beta_\text{acc} &= 0.0209\\
        \gamma &= 0.0390\\
        1 & \text{ fish}
      \end{align*}
    \end{minipage}};
  \end{axis}
\end{tikzpicture}
  \begin{tikzpicture}
  \begin{axis}[
    cycle list name=exotic,
    width=0.48\textwidth, height=0.4\textwidth,
    ybar interval, minor y tick num = 3,
    xmin=-0.5, xmax=10.500000,
    ymin=0, ymax=1,
    xminorgrids=true,
    xtick={0,5,...,40},
    minor x tick num=4,
    x tick label as interval=false,
    bar width=0.4,
    title={Cage 3B autumn 2012 (16 days)},
    xlabel={$n$ -- attached copepod count},
    ylabel={Probability}
    ]
    \addplot +[ybar,bar shift=-0.2] table[x=n,y=p] {data/simulated/sim_3B_2012-10-19.tsv};
    \addlegendentry{Observed}
    \addplot +[ybar,bar shift=0.2] table[x=n,y=q] {data/simulated/sim_3B_2012-10-19.tsv};
    \addlegendentry{Simulated}
    \node at (axis cs:8.000000,0.5) {\begin{minipage}[t]{2cm}
      \small
      \begin{align*}
        \beta_\text{ini} &= 0.0181\\
        \beta_\text{acc} &= 0.0056\\
        \gamma &= 0.0380\\
        45 & \text{ fish}
      \end{align*}
    \end{minipage}};
  \end{axis}
\end{tikzpicture}
\end{center}
\subsection{Cage 4}
\begin{center}
  \begin{tikzpicture}
  \begin{axis}[
    cycle list name=exotic,
    width=0.48\textwidth, height=0.4\textwidth,
    ybar interval, minor y tick num = 3,
    xmin=-0.5, xmax=10.500000,
    ymin=0, ymax=1,
    xminorgrids=true,
    xtick={0,5,...,40},
    minor x tick num=4,
    x tick label as interval=false,
    bar width=0.4,
    title={Cage 4 autumn 2011 (14 days)},
    xlabel={$n$ -- attached copepod count},
    ylabel={Probability}
    ]
    \addplot +[ybar,bar shift=-0.2] table[x=n,y=p] {data/simulated/sim_4_2011-11-09.tsv};
    \addlegendentry{Observed}
    \addplot +[ybar,bar shift=0.2] table[x=n,y=q] {data/simulated/sim_4_2011-11-09.tsv};
    \addlegendentry{Simulated}
    \node at (axis cs:8.000000,0.5) {\begin{minipage}[t]{2cm}
      \small
      \begin{align*}
        \beta_\text{ini} &= 0.0200\\
        \beta_\text{acc} &= 0.0200\\
        \gamma &= 0.0100\\
        48 & \text{ fish}
      \end{align*}
    \end{minipage}};
  \end{axis}
\end{tikzpicture}
  \begin{tikzpicture}
  \begin{axis}[
    cycle list name=exotic,
    width=0.48\textwidth, height=0.4\textwidth,
    ybar interval, minor y tick num = 3,
    xmin=-0.5, xmax=10.500000,
    ymin=0, ymax=1,
    xminorgrids=true,
    xtick={0,5,...,40},
    minor x tick num=4,
    x tick label as interval=false,
    bar width=0.4,
    title={Cage 4 spring 2012 (16 days)},
    xlabel={$n$ -- attached copepod count},
    ylabel={Probability}
    ]
    \addplot +[ybar,bar shift=-0.2] table[x=n,y=p] {data/simulated/sim_4_2012-05-17.tsv};
    \addlegendentry{Observed}
    \addplot +[ybar,bar shift=0.2] table[x=n,y=q] {data/simulated/sim_4_2012-05-17.tsv};
    \addlegendentry{Simulated}
    \node at (axis cs:8.000000,0.5) {\begin{minipage}[t]{2cm}
      \small
      \begin{align*}
        \beta_\text{ini} &= 0.0170\\
        \beta_\text{acc} &= 0.0075\\
        \gamma &= 0.0388\\
        19 & \text{ fish}
      \end{align*}
    \end{minipage}};
  \end{axis}
\end{tikzpicture}\\
  \begin{tikzpicture}
  \begin{axis}[
    cycle list name=exotic,
    width=0.48\textwidth, height=0.4\textwidth,
    ybar interval, minor y tick num = 3,
    xmin=-0.5, xmax=10.500000,
    ymin=0, ymax=1,
    xminorgrids=true,
    xtick={0,5,...,40},
    minor x tick num=4,
    x tick label as interval=false,
    bar width=0.4,
    title={Cage 4 autumn 2012 (16 days)},
    xlabel={$n$ -- attached copepod count},
    ylabel={Probability}
    ]
    \addplot +[ybar,bar shift=-0.2] table[x=n,y=p] {data/simulated/sim_4_2012-10-19.tsv};
    \addlegendentry{Observed}
    \addplot +[ybar,bar shift=0.2] table[x=n,y=q] {data/simulated/sim_4_2012-10-19.tsv};
    \addlegendentry{Simulated}
    \node at (axis cs:8.000000,0.5) {\begin{minipage}[t]{2cm}
      \small
      \begin{align*}
        \beta_\text{ini} &= 0.0153\\
        \beta_\text{acc} &= 0.0095\\
        \gamma &= 0.0394\\
        50 & \text{ fish}
      \end{align*}
    \end{minipage}};
  \end{axis}
\end{tikzpicture}
  \begin{tikzpicture}
  \begin{axis}[
    cycle list name=exotic,
    width=0.48\textwidth, height=0.4\textwidth,
    ybar interval, minor y tick num = 3,
    xmin=-0.5, xmax=10.500000,
    ymin=0, ymax=1,
    xminorgrids=true,
    xtick={0,5,...,40},
    minor x tick num=4,
    x tick label as interval=false,
    bar width=0.4,
    title={Cage 4 autumn 2013 (13 days)},
    xlabel={$n$ -- attached copepod count},
    ylabel={Probability}
    ]
    \addplot +[ybar,bar shift=-0.2] table[x=n,y=p] {data/simulated/sim_4_2013-10-15.tsv};
    \addlegendentry{Observed}
    \addplot +[ybar,bar shift=0.2] table[x=n,y=q] {data/simulated/sim_4_2013-10-15.tsv};
    \addlegendentry{Simulated}
    \node at (axis cs:8.000000,0.5) {\begin{minipage}[t]{2cm}
      \small
      \begin{align*}
        \beta_\text{ini} &= 0.0767\\
        \beta_\text{acc} &= 0.0582\\
        \gamma &= 0.0417\\
        50 & \text{ fish}
      \end{align*}
    \end{minipage}};
  \end{axis}
\end{tikzpicture}
\end{center}
\subsection{Cage 5}
\begin{center}
  \begin{tikzpicture}
  \begin{axis}[
    cycle list name=exotic,
    width=0.48\textwidth, height=0.4\textwidth,
    ybar interval, minor y tick num = 3,
    xmin=-0.5, xmax=10.500000,
    ymin=0, ymax=1,
    xminorgrids=true,
    xtick={0,5,...,40},
    minor x tick num=4,
    x tick label as interval=false,
    bar width=0.4,
    title={Cage 5 spring 2011 (15 days)},
    xlabel={$n$ -- attached copepod count},
    ylabel={Probability}
    ]
    \addplot +[ybar,bar shift=-0.2] table[x=n,y=p] {data/simulated/sim_5_2011-05-27.tsv};
    \addlegendentry{Observed}
    \addplot +[ybar,bar shift=0.2] table[x=n,y=q] {data/simulated/sim_5_2011-05-27.tsv};
    \addlegendentry{Simulated}
    \node at (axis cs:8.000000,0.5) {\begin{minipage}[t]{2cm}
      \small
      \begin{align*}
        \beta_\text{ini} &= 0.0229\\
        \beta_\text{acc} &= 0.0150\\
        \gamma &= 0.0440\\
        47 & \text{ fish}
      \end{align*}
    \end{minipage}};
  \end{axis}
\end{tikzpicture}
  \begin{tikzpicture}
  \begin{axis}[
    cycle list name=exotic,
    width=0.48\textwidth, height=0.4\textwidth,
    ybar interval, minor y tick num = 3,
    xmin=-0.5, xmax=15.500000,
    ymin=0, ymax=1,
    xminorgrids=true,
    xtick={0,5,...,40},
    minor x tick num=4,
    x tick label as interval=false,
    bar width=0.4,
    title={Cage 5 autumn 2011 (14 days)},
    xlabel={$n$ -- attached copepod count},
    ylabel={Probability}
    ]
    \addplot +[ybar,bar shift=-0.2] table[x=n,y=p] {data/simulated/sim_5_2011-11-09.tsv};
    \addlegendentry{Observed}
    \addplot +[ybar,bar shift=0.2] table[x=n,y=q] {data/simulated/sim_5_2011-11-09.tsv};
    \addlegendentry{Simulated}
    \node at (axis cs:12.000000,0.5) {\begin{minipage}[t]{2cm}
      \small
      \begin{align*}
        \beta_\text{ini} &= 0.0200\\
        \beta_\text{acc} &= 0.0200\\
        \gamma &= 0.0100\\
        47 & \text{ fish}
      \end{align*}
    \end{minipage}};
  \end{axis}
\end{tikzpicture}\\
  \begin{tikzpicture}
  \begin{axis}[
    cycle list name=exotic,
    width=0.48\textwidth, height=0.4\textwidth,
    ybar interval, minor y tick num = 3,
    xmin=-0.5, xmax=10.500000,
    ymin=0, ymax=1,
    xminorgrids=true,
    xtick={0,5,...,40},
    minor x tick num=4,
    x tick label as interval=false,
    bar width=0.4,
    title={Cage 5 spring 2012 (16 days)},
    xlabel={$n$ -- attached copepod count},
    ylabel={Probability}
    ]
    \addplot +[ybar,bar shift=-0.2] table[x=n,y=p] {data/simulated/sim_5_2012-05-17.tsv};
    \addlegendentry{Observed}
    \addplot +[ybar,bar shift=0.2] table[x=n,y=q] {data/simulated/sim_5_2012-05-17.tsv};
    \addlegendentry{Simulated}
    \node at (axis cs:8.000000,0.5) {\begin{minipage}[t]{2cm}
      \small
      \begin{align*}
        \beta_\text{ini} &= 0.0154\\
        \beta_\text{acc} &= 0.0094\\
        \gamma &= 0.0394\\
        31 & \text{ fish}
      \end{align*}
    \end{minipage}};
  \end{axis}
\end{tikzpicture}
  \begin{tikzpicture}
  \begin{axis}[
    cycle list name=exotic,
    width=0.48\textwidth, height=0.4\textwidth,
    ybar interval, minor y tick num = 3,
    xmin=-0.5, xmax=10.500000,
    ymin=0, ymax=1,
    xminorgrids=true,
    xtick={0,5,...,40},
    minor x tick num=4,
    x tick label as interval=false,
    bar width=0.4,
    title={Cage 5 autumn 2012 (16 days)},
    xlabel={$n$ -- attached copepod count},
    ylabel={Probability}
    ]
    \addplot +[ybar,bar shift=-0.2] table[x=n,y=p] {data/simulated/sim_5_2012-10-19.tsv};
    \addlegendentry{Observed}
    \addplot +[ybar,bar shift=0.2] table[x=n,y=q] {data/simulated/sim_5_2012-10-19.tsv};
    \addlegendentry{Simulated}
    \node at (axis cs:8.000000,0.5) {\begin{minipage}[t]{2cm}
      \small
      \begin{align*}
        \beta_\text{ini} &= 0.0153\\
        \beta_\text{acc} &= 0.0095\\
        \gamma &= 0.0394\\
        50 & \text{ fish}
      \end{align*}
    \end{minipage}};
  \end{axis}
\end{tikzpicture}\\
  \begin{tikzpicture}
  \begin{axis}[
    cycle list name=exotic,
    width=0.48\textwidth, height=0.4\textwidth,
    ybar interval, minor y tick num = 3,
    xmin=-0.5, xmax=10.500000,
    ymin=0, ymax=1,
    xminorgrids=true,
    xtick={0,5,...,40},
    minor x tick num=4,
    x tick label as interval=false,
    bar width=0.4,
    title={Cage 5 autumn 2013 (13 days)},
    xlabel={$n$ -- attached copepod count},
    ylabel={Probability}
    ]
    \addplot +[ybar,bar shift=-0.2] table[x=n,y=p] {data/simulated/sim_5_2013-10-15.tsv};
    \addlegendentry{Observed}
    \addplot +[ybar,bar shift=0.2] table[x=n,y=q] {data/simulated/sim_5_2013-10-15.tsv};
    \addlegendentry{Simulated}
    \node at (axis cs:8.000000,0.5) {\begin{minipage}[t]{2cm}
      \small
      \begin{align*}
        \beta_\text{ini} &= 0.0898\\
        \beta_\text{acc} &= 0.0981\\
        \gamma &= 0.0911\\
        50 & \text{ fish}
      \end{align*}
    \end{minipage}};
  \end{axis}
\end{tikzpicture}
\end{center}
\subsection{Cage 6}
\begin{center}
  \begin{tikzpicture}
  \begin{axis}[
    cycle list name=exotic,
    width=0.48\textwidth, height=0.4\textwidth,
    ybar interval, minor y tick num = 3,
    xmin=-0.5, xmax=16.500000,
    ymin=0, ymax=1,
    xminorgrids=true,
    xtick={0,5,...,40},
    minor x tick num=4,
    x tick label as interval=false,
    bar width=0.4,
    title={Cage 6 autumn 2011 (14 days)},
    xlabel={$n$ -- attached copepod count},
    ylabel={Probability}
    ]
    \addplot +[ybar,bar shift=-0.2] table[x=n,y=p] {data/simulated/sim_6_2011-11-09.tsv};
    \addlegendentry{Observed}
    \addplot +[ybar,bar shift=0.2] table[x=n,y=q] {data/simulated/sim_6_2011-11-09.tsv};
    \addlegendentry{Simulated}
    \node at (axis cs:12.800000,0.5) {\begin{minipage}[t]{2cm}
      \small
      \begin{align*}
        \beta_\text{ini} &= 0.0043\\
        \beta_\text{acc} &= 0.0333\\
        \gamma &= 0.0054\\
        42 & \text{ fish}
      \end{align*}
    \end{minipage}};
  \end{axis}
\end{tikzpicture}
  \begin{tikzpicture}
  \begin{axis}[
    cycle list name=exotic,
    width=0.48\textwidth, height=0.4\textwidth,
    ybar interval, minor y tick num = 3,
    xmin=-0.5, xmax=10.500000,
    ymin=0, ymax=1,
    xminorgrids=true,
    xtick={0,5,...,40},
    minor x tick num=4,
    x tick label as interval=false,
    bar width=0.4,
    title={Cage 6 spring 2012 (16 days)},
    xlabel={$n$ -- attached copepod count},
    ylabel={Probability}
    ]
    \addplot +[ybar,bar shift=-0.2] table[x=n,y=p] {data/simulated/sim_6_2012-05-17.tsv};
    \addlegendentry{Observed}
    \addplot +[ybar,bar shift=0.2] table[x=n,y=q] {data/simulated/sim_6_2012-05-17.tsv};
    \addlegendentry{Simulated}
    \node at (axis cs:8.000000,0.5) {\begin{minipage}[t]{2cm}
      \small
      \begin{align*}
        \beta_\text{ini} &= 0.0069\\
        \beta_\text{acc} &= 0.0187\\
        \gamma &= 0.0387\\
        25 & \text{ fish}
      \end{align*}
    \end{minipage}};
  \end{axis}
\end{tikzpicture}\\
  \begin{tikzpicture}
  \begin{axis}[
    cycle list name=exotic,
    width=0.48\textwidth, height=0.4\textwidth,
    ybar interval, minor y tick num = 3,
    xmin=-0.5, xmax=10.500000,
    ymin=0, ymax=1,
    xminorgrids=true,
    xtick={0,5,...,40},
    minor x tick num=4,
    x tick label as interval=false,
    bar width=0.4,
    title={Cage 6 autumn 2012 (16 days)},
    xlabel={$n$ -- attached copepod count},
    ylabel={Probability}
    ]
    \addplot +[ybar,bar shift=-0.2] table[x=n,y=p] {data/simulated/sim_6_2012-10-19.tsv};
    \addlegendentry{Observed}
    \addplot +[ybar,bar shift=0.2] table[x=n,y=q] {data/simulated/sim_6_2012-10-19.tsv};
    \addlegendentry{Simulated}
    \node at (axis cs:8.000000,0.5) {\begin{minipage}[t]{2cm}
      \small
      \begin{align*}
        \beta_\text{ini} &= 0.0069\\
        \beta_\text{acc} &= 0.0187\\
        \gamma &= 0.0387\\
        50 & \text{ fish}
      \end{align*}
    \end{minipage}};
  \end{axis}
\end{tikzpicture}
  \begin{tikzpicture}
  \begin{axis}[
    cycle list name=exotic,
    width=0.48\textwidth, height=0.4\textwidth,
    ybar interval, minor y tick num = 3,
    xmin=-0.5, xmax=10.500000,
    ymin=0, ymax=1,
    xminorgrids=true,
    xtick={0,5,...,40},
    minor x tick num=4,
    x tick label as interval=false,
    bar width=0.4,
    title={Cage 6 autumn 2013 (13 days)},
    xlabel={$n$ -- attached copepod count},
    ylabel={Probability}
    ]
    \addplot +[ybar,bar shift=-0.2] table[x=n,y=p] {data/simulated/sim_6_2013-10-15.tsv};
    \addlegendentry{Observed}
    \addplot +[ybar,bar shift=0.2] table[x=n,y=q] {data/simulated/sim_6_2013-10-15.tsv};
    \addlegendentry{Simulated}
    \node at (axis cs:8.000000,0.5) {\begin{minipage}[t]{2cm}
      \small
      \begin{align*}
        \beta_\text{ini} &= 0.0484\\
        \beta_\text{acc} &= 0.0123\\
        \gamma &= 0.0479\\
        33 & \text{ fish}
      \end{align*}
    \end{minipage}};
  \end{axis}
\end{tikzpicture}
\end{center}
\subsection{Cage 8}
\begin{center}
  \begin{tikzpicture}
  \begin{axis}[
    cycle list name=exotic,
    width=0.48\textwidth, height=0.4\textwidth,
    ybar interval, minor y tick num = 3,
    xmin=-0.5, xmax=38.500000,
    ymin=0, ymax=1,
    xminorgrids=true,
    xtick={0,5,...,40},
    minor x tick num=4,
    x tick label as interval=false,
    bar width=0.4,
    title={Cage 8 autumn 2011 (14 days)},
    xlabel={$n$ -- attached copepod count},
    ylabel={Probability}
    ]
    \addplot +[ybar,bar shift=-0.2] table[x=n,y=p] {data/simulated/sim_8_2011-11-09.tsv};
    \addlegendentry{Observed}
    \addplot +[ybar,bar shift=0.2] table[x=n,y=q] {data/simulated/sim_8_2011-11-09.tsv};
    \addlegendentry{Simulated}
    \node at (axis cs:30.400000,0.5) {\begin{minipage}[t]{2cm}
      \small
      \begin{align*}
        \beta_\text{ini} &= 0.0200\\
        \beta_\text{acc} &= 0.0200\\
        \gamma &= 0.0100\\
        50 & \text{ fish}
      \end{align*}
    \end{minipage}};
  \end{axis}
\end{tikzpicture}
  \begin{tikzpicture}
  \begin{axis}[
    cycle list name=exotic,
    width=0.48\textwidth, height=0.4\textwidth,
    ybar interval, minor y tick num = 3,
    xmin=-0.5, xmax=10.500000,
    ymin=0, ymax=1,
    xminorgrids=true,
    xtick={0,5,...,40},
    minor x tick num=4,
    x tick label as interval=false,
    bar width=0.4,
    title={Cage 8 spring 2012 (16 days)},
    xlabel={$n$ -- attached copepod count},
    ylabel={Probability}
    ]
    \addplot +[ybar,bar shift=-0.2] table[x=n,y=p] {data/simulated/sim_8_2012-05-17.tsv};
    \addlegendentry{Observed}
    \addplot +[ybar,bar shift=0.2] table[x=n,y=q] {data/simulated/sim_8_2012-05-17.tsv};
    \addlegendentry{Simulated}
    \node at (axis cs:8.000000,0.5) {\begin{minipage}[t]{2cm}
      \small
      \begin{align*}
        \beta_\text{ini} &= 0.0069\\
        \beta_\text{acc} &= 0.0187\\
        \gamma &= 0.0387\\
        30 & \text{ fish}
      \end{align*}
    \end{minipage}};
  \end{axis}
\end{tikzpicture}\\
  \begin{tikzpicture}
  \begin{axis}[
    cycle list name=exotic,
    width=0.48\textwidth, height=0.4\textwidth,
    ybar interval, minor y tick num = 3,
    xmin=-0.5, xmax=10.500000,
    ymin=0, ymax=1,
    xminorgrids=true,
    xtick={0,5,...,40},
    minor x tick num=4,
    x tick label as interval=false,
    bar width=0.4,
    title={Cage 8 autumn 2012 (16 days)},
    xlabel={$n$ -- attached copepod count},
    ylabel={Probability}
    ]
    \addplot +[ybar,bar shift=-0.2] table[x=n,y=p] {data/simulated/sim_8_2012-10-19.tsv};
    \addlegendentry{Observed}
    \addplot +[ybar,bar shift=0.2] table[x=n,y=q] {data/simulated/sim_8_2012-10-19.tsv};
    \addlegendentry{Simulated}
    \node at (axis cs:8.000000,0.5) {\begin{minipage}[t]{2cm}
      \small
      \begin{align*}
        \beta_\text{ini} &= 0.0153\\
        \beta_\text{acc} &= 0.0095\\
        \gamma &= 0.0394\\
        50 & \text{ fish}
      \end{align*}
    \end{minipage}};
  \end{axis}
\end{tikzpicture}
  \begin{tikzpicture}
  \begin{axis}[
    cycle list name=exotic,
    width=0.48\textwidth, height=0.4\textwidth,
    ybar interval, minor y tick num = 3,
    xmin=-0.5, xmax=10.500000,
    ymin=0, ymax=1,
    xminorgrids=true,
    xtick={0,5,...,40},
    minor x tick num=4,
    x tick label as interval=false,
    bar width=0.4,
    title={Cage 8 autumn 2013 (13 days)},
    xlabel={$n$ -- attached copepod count},
    ylabel={Probability}
    ]
    \addplot +[ybar,bar shift=-0.2] table[x=n,y=p] {data/simulated/sim_8_2013-10-15.tsv};
    \addlegendentry{Observed}
    \addplot +[ybar,bar shift=0.2] table[x=n,y=q] {data/simulated/sim_8_2013-10-15.tsv};
    \addlegendentry{Simulated}
    \node at (axis cs:8.000000,0.5) {\begin{minipage}[t]{2cm}
      \small
      \begin{align*}
        \beta_\text{ini} &= 0.0067\\
        \beta_\text{acc} &= 0.0187\\
        \gamma &= 0.0386\\
        50 & \text{ fish}
      \end{align*}
    \end{minipage}};
  \end{axis}
\end{tikzpicture}
\end{center}
\subsection{Cage 9}
\begin{center}
  \begin{tikzpicture}
  \begin{axis}[
    cycle list name=exotic,
    width=0.48\textwidth, height=0.4\textwidth,
    ybar interval, minor y tick num = 3,
    xmin=-0.5, xmax=10.500000,
    ymin=0, ymax=1,
    xminorgrids=true,
    xtick={0,5,...,40},
    minor x tick num=4,
    x tick label as interval=false,
    bar width=0.4,
    title={Cage 9 autumn 2011 (14 days)},
    xlabel={$n$ -- attached copepod count},
    ylabel={Probability}
    ]
    \addplot +[ybar,bar shift=-0.2] table[x=n,y=p] {data/simulated/sim_9_2011-11-09.tsv};
    \addlegendentry{Observed}
    \addplot +[ybar,bar shift=0.2] table[x=n,y=q] {data/simulated/sim_9_2011-11-09.tsv};
    \addlegendentry{Simulated}
    \node at (axis cs:8.000000,0.5) {\begin{minipage}[t]{2cm}
      \small
      \begin{align*}
        \beta_\text{ini} &= 0.0145\\
        \beta_\text{acc} &= 0.0217\\
        \gamma &= 0.0077\\
        50 & \text{ fish}
      \end{align*}
    \end{minipage}};
  \end{axis}
\end{tikzpicture}
  \begin{tikzpicture}
  \begin{axis}[
    cycle list name=exotic,
    width=0.48\textwidth, height=0.4\textwidth,
    ybar interval, minor y tick num = 3,
    xmin=-0.5, xmax=10.500000,
    ymin=0, ymax=1,
    xminorgrids=true,
    xtick={0,5,...,40},
    minor x tick num=4,
    x tick label as interval=false,
    bar width=0.4,
    title={Cage 9 spring 2012 (16 days)},
    xlabel={$n$ -- attached copepod count},
    ylabel={Probability}
    ]
    \addplot +[ybar,bar shift=-0.2] table[x=n,y=p] {data/simulated/sim_9_2012-05-17.tsv};
    \addlegendentry{Observed}
    \addplot +[ybar,bar shift=0.2] table[x=n,y=q] {data/simulated/sim_9_2012-05-17.tsv};
    \addlegendentry{Simulated}
    \node at (axis cs:8.000000,0.5) {\begin{minipage}[t]{2cm}
      \small
      \begin{align*}
        \beta_\text{ini} &= 0.0170\\
        \beta_\text{acc} &= 0.0075\\
        \gamma &= 0.0388\\
        50 & \text{ fish}
      \end{align*}
    \end{minipage}};
  \end{axis}
\end{tikzpicture}\\
  \begin{tikzpicture}
  \begin{axis}[
    cycle list name=exotic,
    width=0.48\textwidth, height=0.4\textwidth,
    ybar interval, minor y tick num = 3,
    xmin=-0.5, xmax=10.500000,
    ymin=0, ymax=1,
    xminorgrids=true,
    xtick={0,5,...,40},
    minor x tick num=4,
    x tick label as interval=false,
    bar width=0.4,
    title={Cage 9 autumn 2012 (16 days)},
    xlabel={$n$ -- attached copepod count},
    ylabel={Probability}
    ]
    \addplot +[ybar,bar shift=-0.2] table[x=n,y=p] {data/simulated/sim_9_2012-10-19.tsv};
    \addlegendentry{Observed}
    \addplot +[ybar,bar shift=0.2] table[x=n,y=q] {data/simulated/sim_9_2012-10-19.tsv};
    \addlegendentry{Simulated}
    \node at (axis cs:8.000000,0.5) {\begin{minipage}[t]{2cm}
      \small
      \begin{align*}
        \beta_\text{ini} &= 0.0214\\
        \beta_\text{acc} &= 0.0372\\
        \gamma &= 0.0535\\
        50 & \text{ fish}
      \end{align*}
    \end{minipage}};
  \end{axis}
\end{tikzpicture}
  \begin{tikzpicture}
  \begin{axis}[
    cycle list name=exotic,
    width=0.48\textwidth, height=0.4\textwidth,
    ybar interval, minor y tick num = 3,
    xmin=-0.5, xmax=10.500000,
    ymin=0, ymax=1,
    xminorgrids=true,
    xtick={0,5,...,40},
    minor x tick num=4,
    x tick label as interval=false,
    bar width=0.4,
    title={Cage 9 autumn 2013 (13 days)},
    xlabel={$n$ -- attached copepod count},
    ylabel={Probability}
    ]
    \addplot +[ybar,bar shift=-0.2] table[x=n,y=p] {data/simulated/sim_9_2013-10-15.tsv};
    \addlegendentry{Observed}
    \addplot +[ybar,bar shift=0.2] table[x=n,y=q] {data/simulated/sim_9_2013-10-15.tsv};
    \addlegendentry{Simulated}
    \node at (axis cs:8.000000,0.5) {\begin{minipage}[t]{2cm}
      \small
      \begin{align*}
        \beta_\text{ini} &= 0.0200\\
        \beta_\text{acc} &= 0.0200\\
        \gamma &= 0.0100\\
        50 & \text{ fish}
      \end{align*}
    \end{minipage}};
  \end{axis}
\end{tikzpicture}
\end{center}
\subsection{Cage 10}
\begin{center}
  \begin{tikzpicture}
  \begin{axis}[
    cycle list name=exotic,
    width=0.48\textwidth, height=0.4\textwidth,
    ybar interval, minor y tick num = 3,
    xmin=-0.5, xmax=29.500000,
    ymin=0, ymax=1,
    xminorgrids=true,
    xtick={0,5,...,40},
    minor x tick num=4,
    x tick label as interval=false,
    bar width=0.4,
    title={Cage 10 autumn 2011 (14 days)},
    xlabel={$n$ -- attached copepod count},
    ylabel={Probability}
    ]
    \addplot +[ybar,bar shift=-0.2] table[x=n,y=p] {data/simulated/sim_10_2011-11-09.tsv};
    \addlegendentry{Observed}
    \addplot +[ybar,bar shift=0.2] table[x=n,y=q] {data/simulated/sim_10_2011-11-09.tsv};
    \addlegendentry{Simulated}
    \node at (axis cs:23.200000,0.5) {\begin{minipage}[t]{2cm}
      \small
      \begin{align*}
        \beta_\text{ini} &= 0.0200\\
        \beta_\text{acc} &= 0.0200\\
        \gamma &= 0.0100\\
        50 & \text{ fish}
      \end{align*}
    \end{minipage}};
  \end{axis}
\end{tikzpicture}
  \begin{tikzpicture}
  \begin{axis}[
    cycle list name=exotic,
    width=0.48\textwidth, height=0.4\textwidth,
    ybar interval, minor y tick num = 3,
    xmin=-0.5, xmax=10.500000,
    ymin=0, ymax=1,
    xminorgrids=true,
    xtick={0,5,...,40},
    minor x tick num=4,
    x tick label as interval=false,
    bar width=0.4,
    title={Cage 10 spring 2012 (16 days)},
    xlabel={$n$ -- attached copepod count},
    ylabel={Probability}
    ]
    \addplot +[ybar,bar shift=-0.2] table[x=n,y=p] {data/simulated/sim_10_2012-05-17.tsv};
    \addlegendentry{Observed}
    \addplot +[ybar,bar shift=0.2] table[x=n,y=q] {data/simulated/sim_10_2012-05-17.tsv};
    \addlegendentry{Simulated}
    \node at (axis cs:8.000000,0.5) {\begin{minipage}[t]{2cm}
      \small
      \begin{align*}
        \beta_\text{ini} &= 0.0069\\
        \beta_\text{acc} &= 0.0187\\
        \gamma &= 0.0387\\
        5 & \text{ fish}
      \end{align*}
    \end{minipage}};
  \end{axis}
\end{tikzpicture}\\
  \begin{tikzpicture}
  \begin{axis}[
    cycle list name=exotic,
    width=0.48\textwidth, height=0.4\textwidth,
    ybar interval, minor y tick num = 3,
    xmin=-0.5, xmax=10.500000,
    ymin=0, ymax=1,
    xminorgrids=true,
    xtick={0,5,...,40},
    minor x tick num=4,
    x tick label as interval=false,
    bar width=0.4,
    title={Cage 10 autumn 2012 (16 days)},
    xlabel={$n$ -- attached copepod count},
    ylabel={Probability}
    ]
    \addplot +[ybar,bar shift=-0.2] table[x=n,y=p] {data/simulated/sim_10_2012-10-19.tsv};
    \addlegendentry{Observed}
    \addplot +[ybar,bar shift=0.2] table[x=n,y=q] {data/simulated/sim_10_2012-10-19.tsv};
    \addlegendentry{Simulated}
    \node at (axis cs:8.000000,0.5) {\begin{minipage}[t]{2cm}
      \small
      \begin{align*}
        \beta_\text{ini} &= 0.0153\\
        \beta_\text{acc} &= 0.0095\\
        \gamma &= 0.0394\\
        50 & \text{ fish}
      \end{align*}
    \end{minipage}};
  \end{axis}
\end{tikzpicture}
  \begin{tikzpicture}
  \begin{axis}[
    cycle list name=exotic,
    width=0.48\textwidth, height=0.4\textwidth,
    ybar interval, minor y tick num = 3,
    xmin=-0.5, xmax=10.500000,
    ymin=0, ymax=1,
    xminorgrids=true,
    xtick={0,5,...,40},
    minor x tick num=4,
    x tick label as interval=false,
    bar width=0.4,
    title={Cage 10 autumn 2013 (13 days)},
    xlabel={$n$ -- attached copepod count},
    ylabel={Probability}
    ]
    \addplot +[ybar,bar shift=-0.2] table[x=n,y=p] {data/simulated/sim_10_2013-10-15.tsv};
    \addlegendentry{Observed}
    \addplot +[ybar,bar shift=0.2] table[x=n,y=q] {data/simulated/sim_10_2013-10-15.tsv};
    \addlegendentry{Simulated}
    \node at (axis cs:8.000000,0.5) {\begin{minipage}[t]{2cm}
      \small
      \begin{align*}
        \beta_\text{ini} &= 0.0067\\
        \beta_\text{acc} &= 0.0187\\
        \gamma &= 0.0386\\
        50 & \text{ fish}
      \end{align*}
    \end{minipage}};
  \end{axis}
\end{tikzpicture}
\end{center}

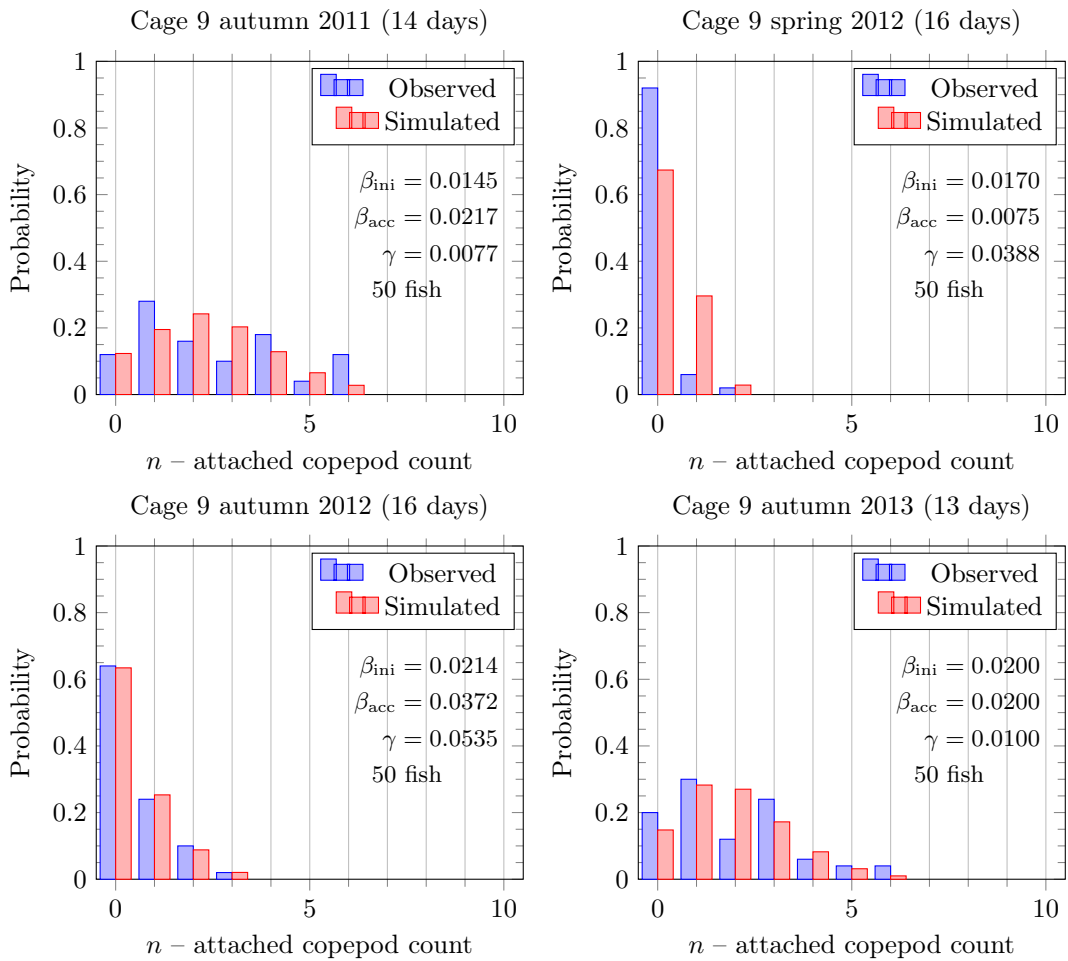
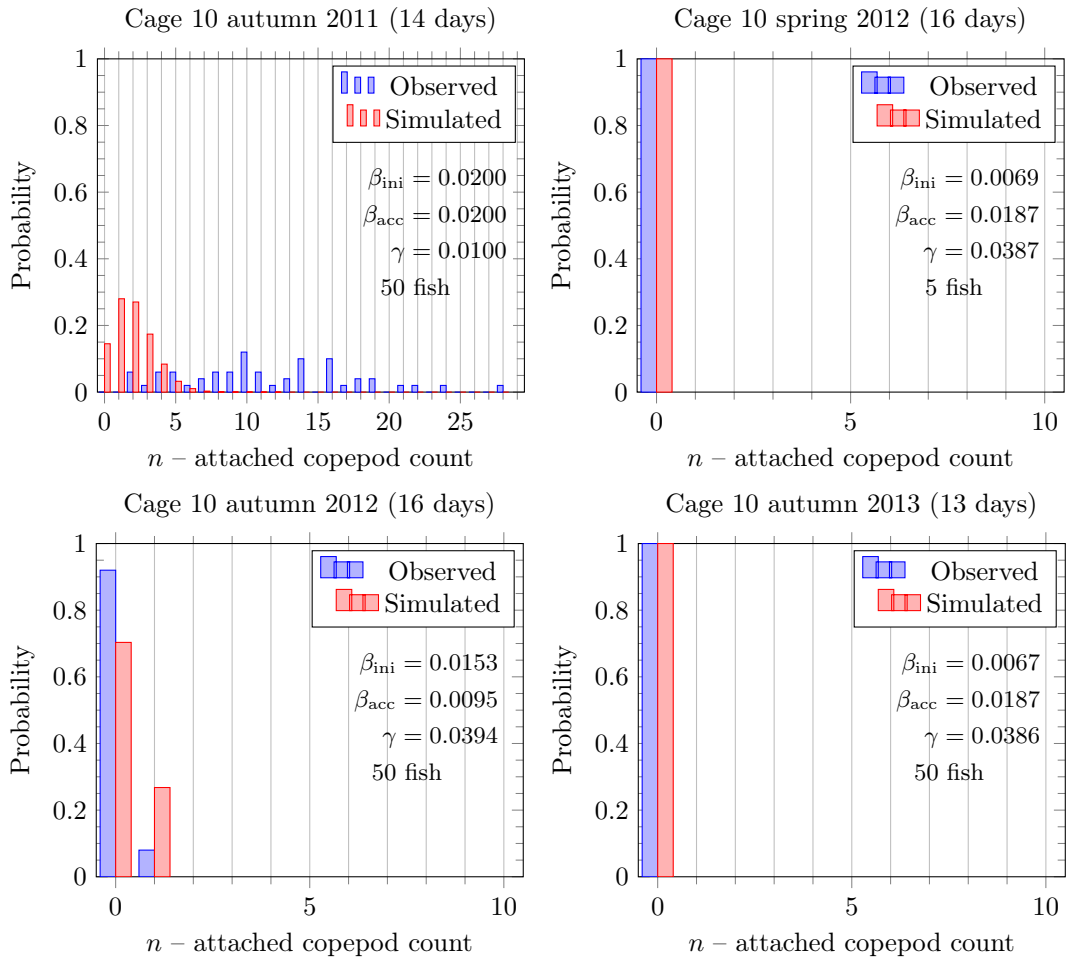
\end{document}